\documentclass[aps,
twocolumn,superscriptaddress,showpacs,showkeys,floatfix]{revtex4-1}
\usepackage{graphicx}
\usepackage{charter}
\usepackage{amssymb}
\usepackage{setspace}
\usepackage{amsmath}
\usepackage{csquotes}
\usepackage{float}
\usepackage {braket}
\usepackage{hyperref}
\usepackage{xcolor}
\usepackage{xfrac}
\usepackage{siunitx}
\usepackage{tabularx}
\usepackage{tabulary}
\usepackage{tabu}
\usepackage{mathtools}
\usepackage{xspace}
\usepackage{booktabs}
\usepackage{titlesec}
\titlespacing{\section}
{0pt}{1.5ex plus 0.2ex minus .2ex}{1.ex  plus 0.5ex minus .5ex}

\definecolor{dark-red}{rgb}{0.4,0.15,0.15}
\definecolor{dark-blue}{rgb}{0.15,0.15,0.4}
\definecolor{medium-blue}{rgb}{0,0,0.5}
\hypersetup{
    colorlinks, linkcolor={dark-red},
    citecolor={dark-blue}, urlcolor={medium-blue}
}

\newcommand{\RuCl}{$\alpha$-RuCl$_3$\xspace}
\newcommand{\musr}{$\mu^+$SR\xspace}

\begin{document}


\title{Novel magnetism on a honeycomb lattice in \RuCl studied by muon spin rotation}

\author{F.~Lang}
\email{franz.lang@physics.ox.ac.uk}
\affiliation{Oxford University Department of Physics, Clarendon Laboratory, Parks Road, Oxford, OX1 3PU, United Kingdom}

\author{ P.~J.~Baker}
\affiliation{ISIS Facility, Rutherford Appleton Laboratory, Chilton, Oxfordshire OX11 0QX, United Kingdom}

\author{ A.~A.~Haghighirad}
\affiliation{Oxford University Department of Physics, Clarendon Laboratory, Parks Road, Oxford, OX1 3PU, United Kingdom}

\author{Y.~Li}
\affiliation{Institut f\"{u}r Theoretische Physik, Goethe-Universit\"{a}t Frankfurt, 60438 Frankfurt am Main, Germany}

\author{D.~Prabhakaran}
\affiliation{Oxford University Department of Physics, Clarendon Laboratory, Parks Road, Oxford, OX1 3PU, United Kingdom}

\author{R.~Valent\'{\i}}
\affiliation{Institut f\"{u}r Theoretische Physik, Goethe-Universit\"{a}t Frankfurt, 60438 Frankfurt am Main, Germany}

\author{S.~J.~Blundell}
\email{s.blundell@physics.ox.ac.uk}
\affiliation{Oxford University Department of Physics, Clarendon Laboratory, Parks Road, Oxford, OX1 3PU, United Kingdom}

\date{\today}

\begin{abstract}
Muon spin rotation measurements have been performed on a powder sample
of \RuCl, a layered material which previously has been proposed to be a
quantum magnet on a honeycomb lattice close to a quantum spin liquid
ground state. Our data reveal two distinct phase transitions at 11\,K and 14\,K which we interpret as originating from the onset of three-dimensional order and in-plane magnetic order, respectively.
We identify, with the help of density functional theory calculations, likely muon stopping sites and combine these with
dipolar field calculations to show that the two measured muon rotation
frequencies are consistent with two inequivalent muon sites within a
zig-zag antiferromagnetic structure proposed previously.  
\end{abstract}


\maketitle

Solid-state systems with architectures that contain triangles or tetrahedra offer the possibility of realizing novel magnetically frustrated states, such as quantum spin liquids~\cite{Balents2010} or exotic topological phases~\cite{Jackeli2009}. 
One such candidate system for frustrated magnetism is \RuCl, which adopts the honeycomb structure. It is thought to be a spin-orbit assisted Mott insulator~\cite{Zhou2016,Pollini1994}, in which both the near two-dimensionality of the separate honeycomb layers and bond-dependent interactions, which may embody Kitaev physics, are proposed to be major ingredients~\cite{Plumb2014}. Unconventional excitations observed via Raman~\cite{Sandilands2015} and inelastic neutron scattering~\cite{Banerjee2016} have been presented as evidence that \RuCl may be close to a quantum spin liquid ground state.
Various magnetic transitions have been reported in \RuCl with early studies pointing towards an antiferromagnetic transition with numerous reported temperatures of $13$~K~\cite{Fletcher1967}, $15.6$~K~\cite{Kobayashi1992} or even $30$~K~\cite{Fletcher1963}, while later investigations proposed a potential second transition around $8$~K~\cite{Majumder2015, Sears2015,Kubota2015} thought to originate from low-moment magnetism. Recent neutron powder diffraction provided evidence for a single transition to a zig-zag antiferromagnetic state with 2-layer stacking at $T_\text{N}\!=\!13$~K~\cite{Johnson2015}, though a later single crystal neutron study has proposed a single transition at $8$~K to a 3-layer stacking magnetic order in pristine single crystals and a change of $T_\text{c}$ to $14$~K upon mechanical deformation of the crystals~\cite{Cao2016}. These differences in observed properties could be due to the propensity of this compound to exhibit stacking faults between the weakly coupled honeycomb layers~\cite{Johnson2015}.

Positive muons as local magnetic probes present an ideal tool for
detecting magnetic order and characterizing magnetic behavior, and
have been extensively utilized in muon-spin rotation or relaxation
(\musr) studies of frustrated systems~\cite{Caretta2011}. Here, we
present results from zero-field (ZF) \musr investigations of \RuCl
powder complemented by a theoretical analysis based on density
functional theory (DFT) and dipolar field calculations. Below about
$14$~K our sample shows clear evidence for long-range magnetic order,
with two muon precession signals resolvable at low temperature. 
However, there are clear indications of the higher frequency signal vanishing at a slightly lower temperature of about $11$~K.

Polycrystalline samples of \RuCl were synthesized by vacuum sublimation from commercial RuCl$_3$ powder (Sigma Aldrich), which was sealed in a quartz ampoule ($p\!\approx\! 10^{-5}$~mbar) and placed in a three-zone furnace with a hot and cold end of $\SI{650}{\degreeCelsius}$ and $\SI{450}{\degreeCelsius}$, respectively. Those temperatures were chosen in order to obtain phase-pure \RuCl (the $\beta$-polytype transforms irreversibly into the $\alpha$-phase above $\SI{395}{\degreeCelsius}$) and to keep the Cl$_2$ gas pressure in the ampoule below atmospheric pressure. The polycrystalline material harvested from the ampoule contained many plate-like shiny crystals of hexagonal shape. X-ray diffraction confirmed the samples to be single phase and in agreement with the $C2/m$ structure~\cite{Johnson2015,Cao2016}. See the Supplemental Material~\cite{SuppMat} for more details on the X-ray characterization.

We conducted ZF \musr measurements of a powder sample of \RuCl on the EMU spectrometer at the ISIS muon facility, RAL (UK), as well as the GPS spectrometer at the Swiss Muon Source, PSI (Switzerland). Data were collected in the temperature range $1.5$~K to $40$~K using $^4$He cryostats.
In a \musr experiment spin-polarized muons are implanted into a
sample, where they Larmor-precess around the local magnetic field at
the muon stopping site.  By measuring the angular distribution of the
decay product positrons the spin polarization can be tracked. 
In the case of long-range magnetic order, coherent magnetic fields at particular muon stopping sites within the unit cell lead to oscillatory signals with frequencies dependent on the local magnetic fields at each site. In \musr impurity phases only contribute
according to their volume fraction, and so the technique is an
effective measure of intrinsic behavior.

\begin{figure}[htb] \center
\includegraphics[width=\columnwidth, clip, trim= 0.0mm 0.0mm 0.0mm 0.0mm]{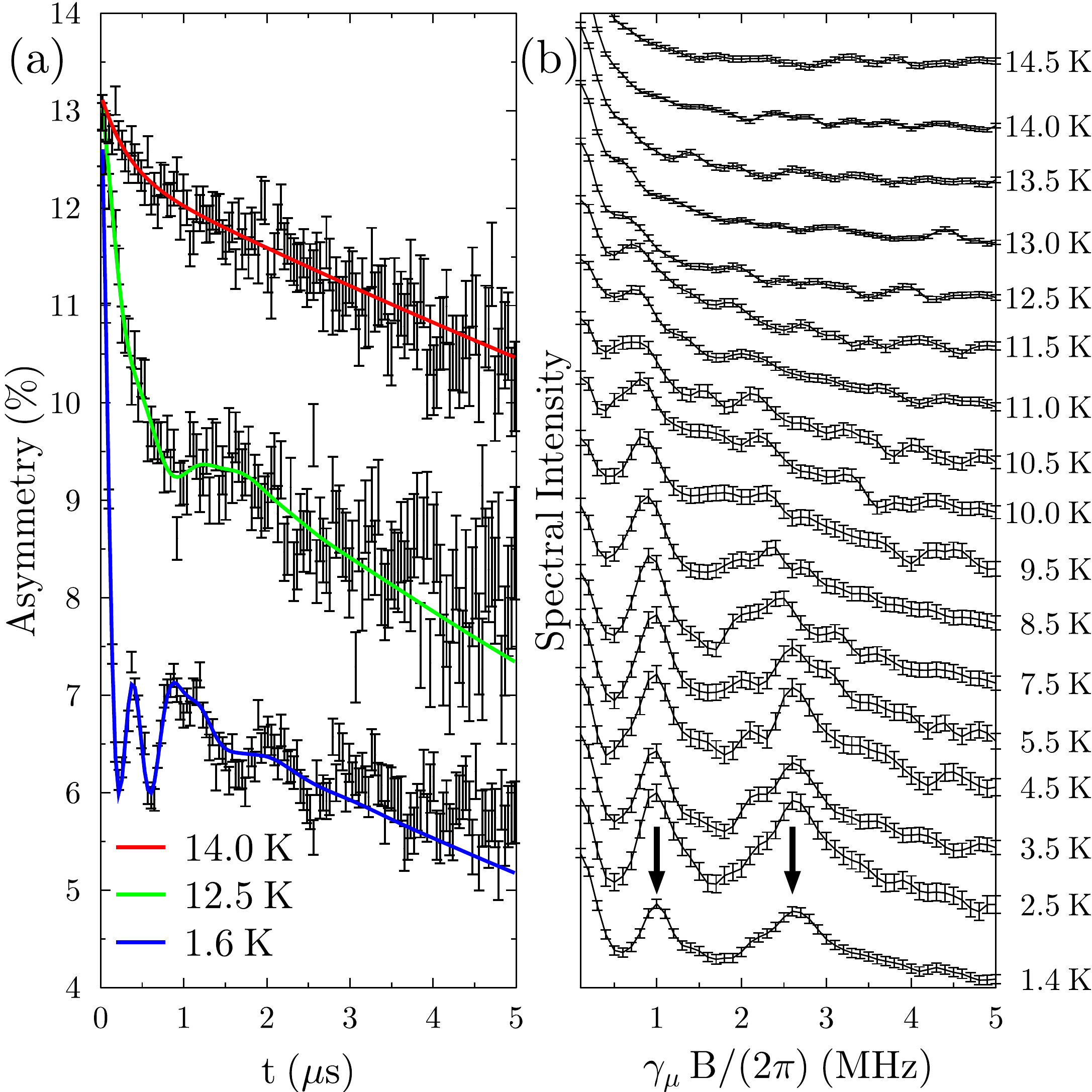}
\caption{ \label{plot:asymsfft}
Panel (a): Muon asymmetries at selected temperatures. Solid lines represent fits using two ($T\!\lesssim\!
11$~K) or one (11~K$\lesssim\! T\! \lesssim\! 14$~K) oscillating components with a Lorentzian relaxation. Panel (b): Fourier transform spectra of the muon asymmetries (vertically displaced for clarity).
}
\end{figure}

Representative raw data obtained are plotted in Fig.~\ref{plot:asymsfft}(a) with Fourier transform spectra presented in Fig.~\ref{plot:asymsfft}(b). The measurements reveal oscillations below $14$~K with two clearly separate frequencies at low temperatures around $1$~MHz and $2.5$~MHz, resulting from two inequivalent muon stopping sites with local fields of $7.5$~mT and $18.5$~mT, respectively.

\begin{figure}[htb] 
\center
\includegraphics[width=0.9\columnwidth, clip, trim= 0.5mm 5mm 0.5mm 2.3mm]{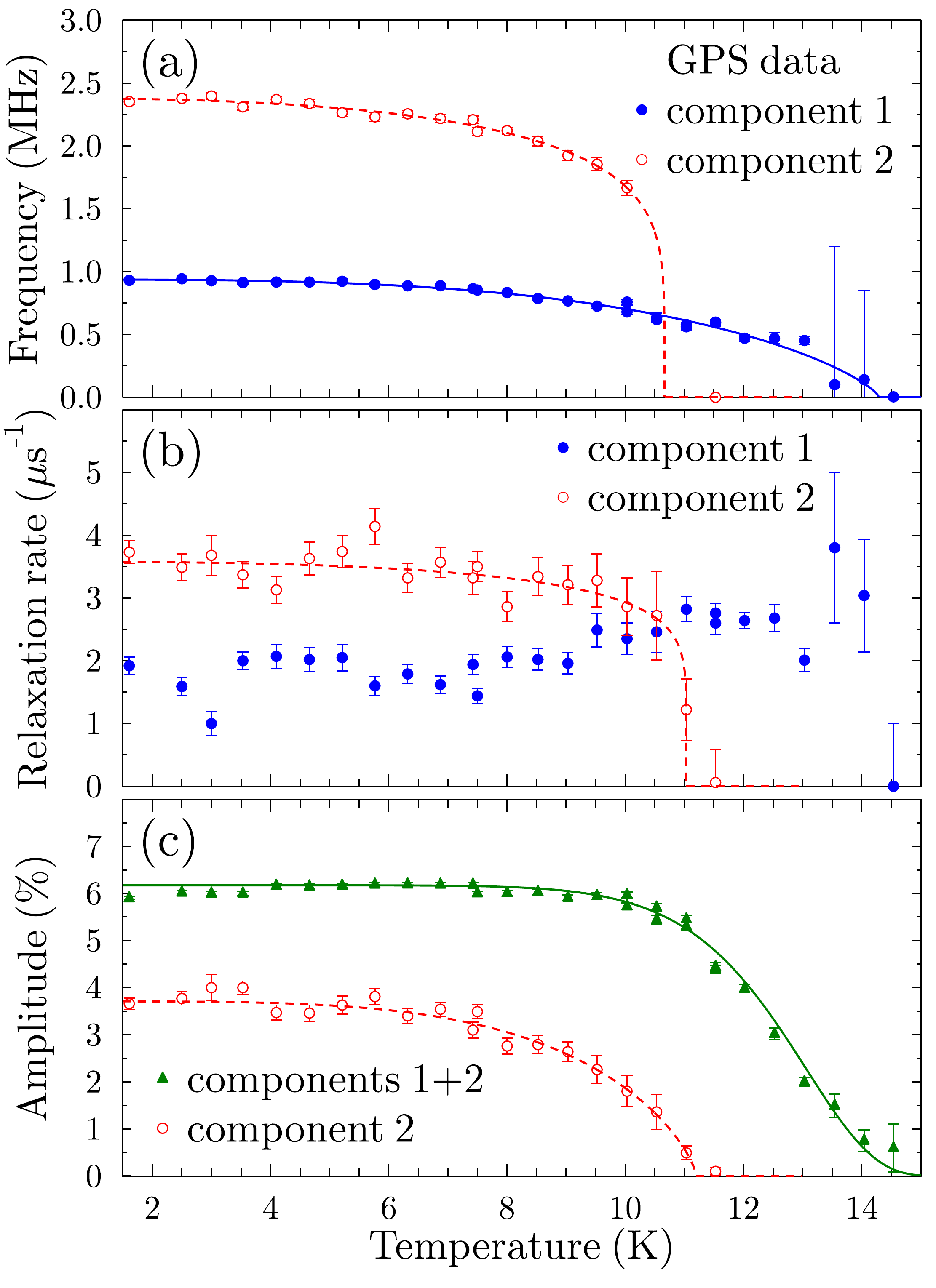}
\caption{ \label{plot:GPSEMUfits}
Results of fitting two oscillation frequencies with Lorentzian relaxation to the muon asymmetry.  The lines represent order parameter fits of the form $y^2=y_0^2(1-(x/T_c)^\alpha)^\beta+c^2$.
}
\end{figure}

The \musr data can be well fitted below $11$~K with a sum of two oscillating functions $\cos\omega_it$ multiplied by exponentials of the form $e^{-\lambda_i t}$, allowing for relaxation caused by slow dynamics of the magnetic moments. In the range $11\text{~K}\!\lesssim\! T \!\lesssim\! 14 \text{~K}$ only one such oscillating component is required. Figure~\ref{plot:GPSEMUfits} presents the resulting frequencies $\omega_i$, relaxation rates $\lambda_i$ and oscillation amplitudes of the precession signals for the data collected on the GPS spectrometer. Essentially identical results were obtained in a separate experiment using the EMU spectrometer, demonstrating reproducibility. The fitted parameters can be modeled with a phenomenological order parameter equation of the form  $y^2=y_0^2(1-(x/T_c)^\alpha)^\beta+c^2$ to give critical temperatures of $11.0(5)$~K and $14.3(3)$~K for the high and low frequency components, respectively. 
The presence of two \musr precession signals necessitates two inequivalent muon stopping sites in the magnetic phase of our sample, whose origin we discuss later.

Further analysis requires the knowledge of the potential muon stopping sites. Therefore, we employ DFT calculations to map out the electrostatic Coulomb potential of \RuCl throughout its unit cell. The maxima of such a potential map have been a reliable approximation to the muon sites in previous more in-depth ``DFT+$\mu$'' calculations, which also accounted for local distortions of the lattice caused by the muon presence~\cite{Moller2013_DFT,Foronda2015,johannesthesis}. 

We performed DFT calculations within the 
generalized gradient approximation~\cite{Perdew1996} by employing the 
full potential linearized augmented plane wave (LAPW) basis as implemented 
in  WIEN2k~\cite{Wien2k}. The $RK_{max}$ parameter was set to
 9 and we used a mesh of 800 ${\bf k}$ points in the first Brillouin zone.
The electrostatic (Coulomb) potential was calculated
from the converged electron density and the three-dimensional
electrostatic potential maps were obtained with the XCrySDen package~\cite{Kokalj1999} and visualized with the Vesta software~\cite{Momma2008}.

\begin{figure}[htb] 
\center
\includegraphics[width=\columnwidth, clip, trim= 0.0mm 1.0mm 0.0mm 1.0mm]{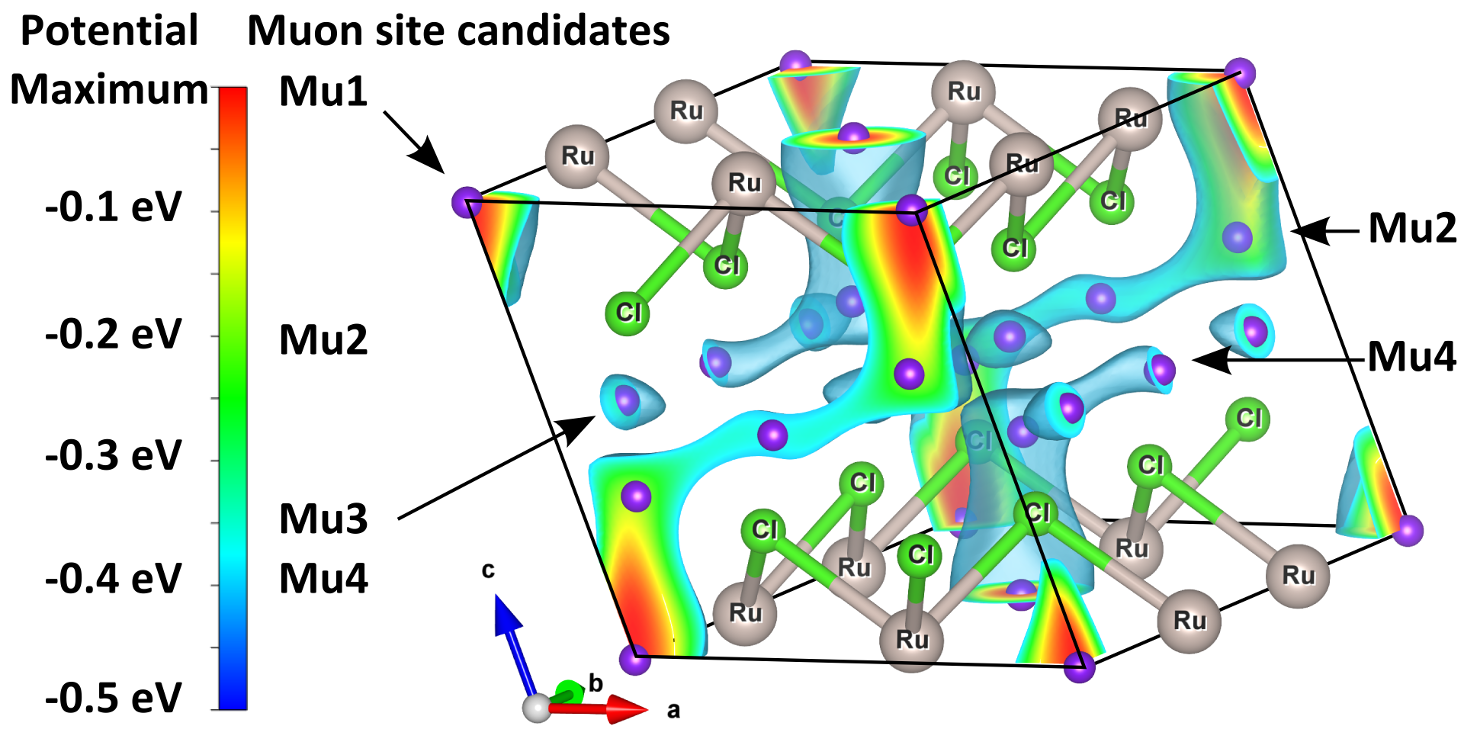}
\caption{ \label{plot:coulomb}
Coulomb potential of \RuCl calculated via DFT. The blue isosurface plotted is at $0.4$~eV below the maximum. The purple spheres indicate the muon site candidates we identified. Their labels are placed next to the color scale to indicate the approximate value of the potential at the sites.
}
\end{figure}

The Coulomb potential of \RuCl calculated via DFT is plotted in
Figure~\ref{plot:coulomb}, with the global maximum of the potential
chosen as the reference value. A large Coulomb potential corresponds
to a low energy required to add a positive charge. Therefore, by
considering regions of high electrostatic potential, and particularly
local maxima, we can identify plausible regions for a muon to stop
in. When additionally taking into account that we expect a $\mu^+$ to
implant near a Cl$^-$ ion~\cite{SuppMat}, we find four plausible muon site
candidates, which are shown in Figure~\ref{plot:coulomb} and
summarized in Table~\ref{tab:coulomb}. These candidate sites are
separated by up to $0.4$~eV in their Coulomb potential values, with
the origin (Mu1) being the lowest.
While the muon will generally perturb its local environment, its
effect is short-ranged and significant only for the nearest neighbor
ions~\cite{Moller2013_DFT, Foronda2015}, and in the present case we
anticipate only a small displacement of a nearest Cl$^-$ ion and
negligible effect on the magnetic moment carrying Ru$^{3+}$ ions. As a
result, we do not expect distortions to have a significant impact on
the bulk magnetism probed in our \musr measurement.

\begin{table}[htb]
\begin{tabular*}{\columnwidth}{@{}l @{\extracolsep{\fill}}*5c @{\extracolsep{\fill}}c@{}}
Atom & WP& SS & x & y & z \\
 \hline
 Ru & 4g & 2 & 0 & 0.33441 & 0\\ 
 Cl & 4i & m & 0.73023 & 0 &0.23895\\ 
 Cl & 8j & 1 & 0.75138 & 0.17350 & 0.76619\\ 
\hline
 Mu1 & 2a & 2/m & 0 & 0 & 0\\ 
 Mu2 & 4i & m & 0.14 &0 & 0.36\\ 
 Mu3 & 4g & 2 & 0 & 0.2 & 0.5\\ 
 Mu4 & 2d & 2/m & 0.5 & 0 & 0.5\\ 
\hline 
\end{tabular*}
\caption{ \label{tab:coulomb}
Fractional coordinates of atoms and muon site candidates determined through DFT calculations. Abbreviations stand for Wyckoff position (WP) and site symmetry (SS). The fractional coordinates of \RuCl originate from Ref.~\cite{Johnson2015} and are compatible with x-ray diffraction characterization~\cite{SuppMat}.
}
\end{table}

We now calculate the local magnetic field experienced by an implanted
muon. This field is in general a sum of contributions due to dipolar
couplings, demagnetizing and Lorentz fields and hyperfine
interactions. Since \RuCl orders antiferromagnetically the
demagnetizing and Lorentz fields are zero. We expect the $\mu^+$ to stop near Cl$^-$ ions and thus direct overlap with any Ru$^{3+}$ electron spin density will be tiny and so we neglect any hyperfine contribution~\cite{Moller2013_DFT,Moller2013_FuF}. Therefore, we focus on the dominant dipole field only, which for a muon at position $\boldsymbol{r}_\mu$ and magnetic moments $\boldsymbol{\mu}_i$ at $\boldsymbol{r}_i$ is given by
\begin{equation}
\boldsymbol{{B}}_\mathrm{dip}(\boldsymbol{r}_\mu)=\sum_i \frac{\mu_o}{4\pi |\Delta\boldsymbol{r}_i|^3}\left[\frac{3(\boldsymbol{\mu}_i\cdot\Delta\boldsymbol{r}_i)\Delta\boldsymbol{r}_i}{|\Delta\boldsymbol{r}_i|^2}-\boldsymbol{\mu}_i\right]
,\end{equation}
where $\Delta\boldsymbol{r}_i=\boldsymbol{r}_i-\boldsymbol{r}_\mu$.

There exists substantial knowledge about the magnetic structure of \RuCl based on neutron diffraction experiments. One neutron powder study provided evidence for a zig-zag antiferromagnetic order within each Ru honeycomb layer with an additional antiferromagnetic stacking between the layers. The corresponding propagation vector is $\boldsymbol{k}=(0,1,0.5)$, and moreover the moments are constrained to lie in the $ac$ plane and the lower limit of the moment size is $0.64(4)\mu_\mathrm{B}$~\cite{Johnson2015}. However, another recent single crystal measurement proposed an alternative zig-zag antiferromagnetic ordering with 3-layer stacking ($\boldsymbol{k}=(0,1,1/3)$) in pristine single crystals with moments aligning in the $ac$ plane in a spiral or collinear pattern~\cite{Cao2016}. 
Investigations using ab initio and model calculations also find an in-plane zig-zag antiferromagnetic order~\cite{Winter2016,Kim2015} and predict the magnetic moments to make an angle of $\approx\!\SI{30}{\degree}$ with the $ab$ plane~\cite{Winter2016,Chaloupka2015}. 

\begin{figure}[htb] 
\center
\includegraphics[width=\columnwidth, clip, trim= 7.0mm 10.0mm 12.0mm 1.0mm]{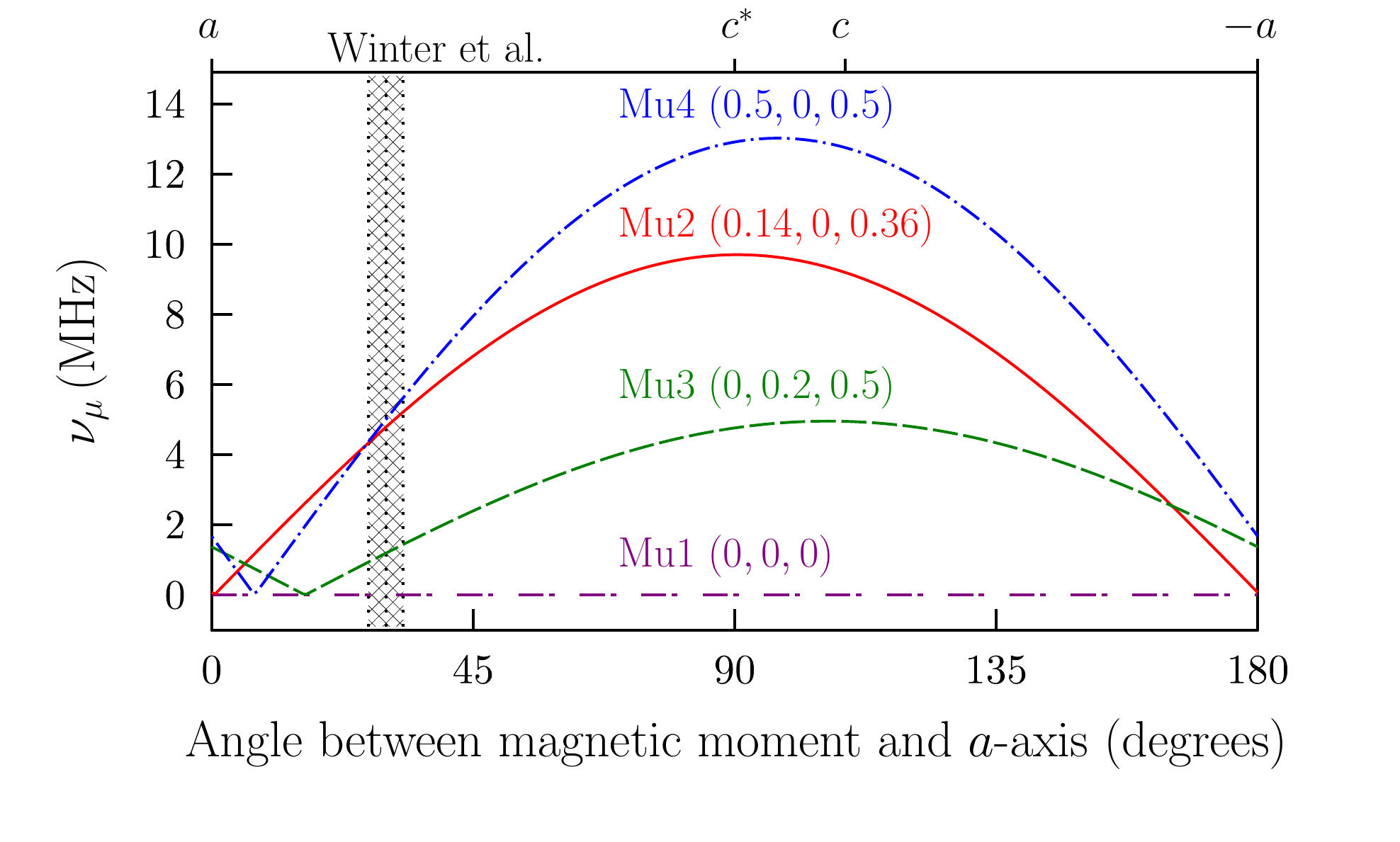}
\caption{ \label{plot:dipoleangle}
Muon Larmor precession frequencies due to dipolar fields at the four muon site candidates as a function of the magnetic moment direction in the $ac$-plane. Directions parallel to crystallographic axes are indicated at the top of the plot. The magnetic structure was taken to be the 2-layer ordering proposed by Johnson {\it et al.}~\cite{Johnson2015} and the approximate moment direction predicted by Winter {\it et al.}~\cite{Winter2016} has been marked.
}
\end{figure}

\begin{figure}[htb] 
\center
\includegraphics[width=\columnwidth, clip, trim= 4.0mm 8.0mm 13.0mm 9.0mm]{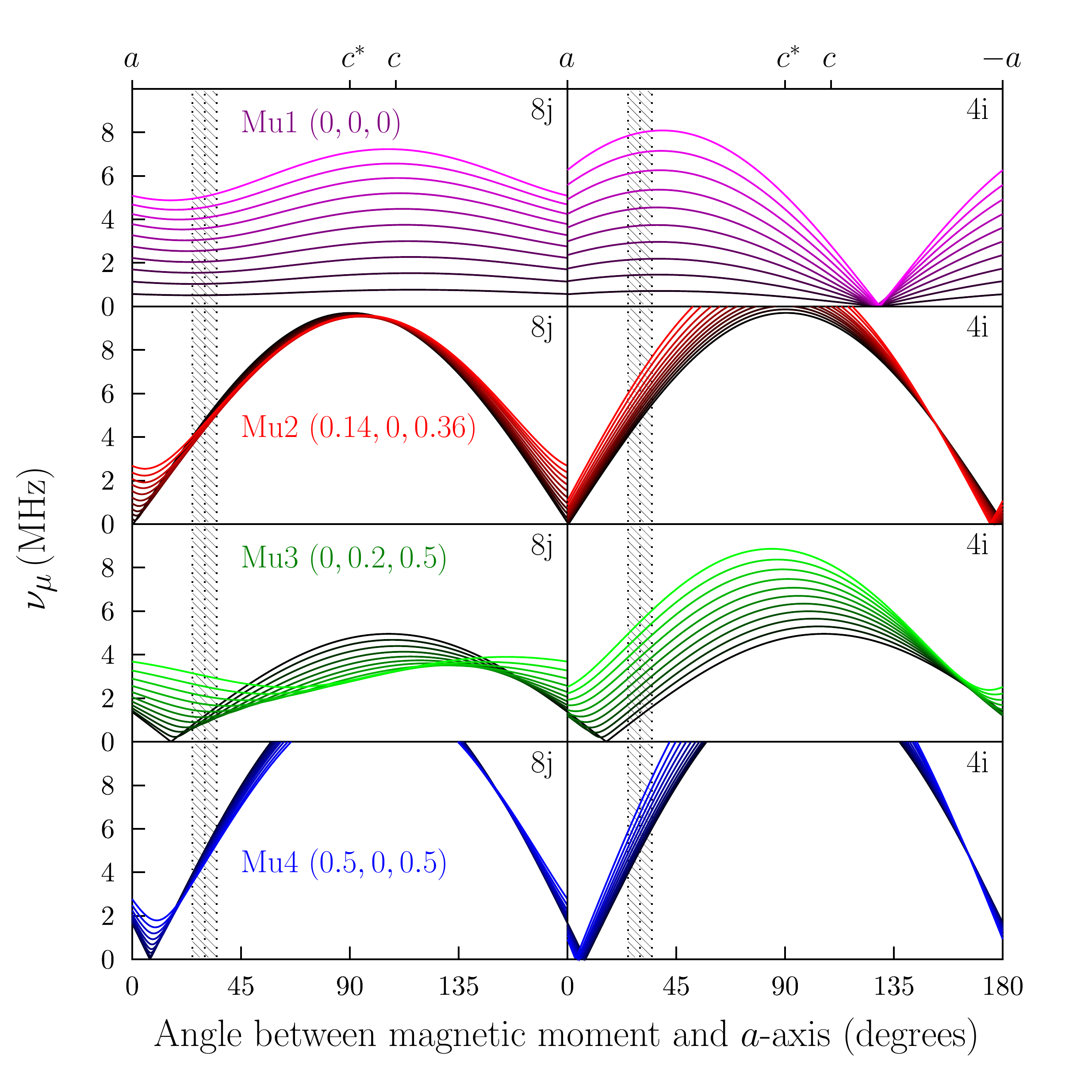}
\caption{ \label{plot:dipoledistort}
Muon Larmor precession frequencies at the muon site candidates (black curves) and for ten positions ($\SI{0.1}{\angstrom}$ between each) along a straight line towards the nearest Cl$^-$ ions (coloured curves), as a function of the magnetic moment direction in the $ac$ plane. Positions further away from the undistorted muon site candidates are displayed more colourfully. The left and right columns show distortions towards Cl$^-$ ions on the $8j$ and $4i$ Wyckoff positions, respectively (see Table~\ref{tab:coulomb}). The magnetic structure was taken to be the 2-layer ordering proposed by Johnson {\it et al.}~\cite{Johnson2015}. Moment directions parallel to crystallographic axes are indicated at the top and the approximate angle predicted by Winter {\it et al.}~\cite{Winter2016} is marked by the dotted vertical lines.
}
\end{figure}

Using the known crystal structure and the proposed 2-layer magnetic
ordering we computed the dipole field strength at the candidate muon
sites obtained through DFT simulations. Figure~\ref{plot:dipoleangle}
displays the resulting Larmor frequencies and how they change as a
function of the magnetic moment direction within the $ac$ plane. Note
that the dipole field vanishes due to the local symmetry at candidate
site Mu1, which is the electrostatically most favourable
one. Figure~\ref{plot:dipoleangle} reveals that there is no single
moment direction within the $ac$ plane for which we obtain precession
frequencies that agree with both the experimentally observed ones. We
can improve our estimates by incorporating the fact that we expect the
muon to form a bond with a nearby Cl$^-$ ion of length
$\approx\!\SI{1.5}{\angstrom}$~\cite{SuppMat}. Our revised model considers the muon
site to be displaced from our earlier candidate sites towards each of
the nearest Cl$^-$ ions. Figure~\ref{plot:dipoledistort} presents the
resulting muon precession frequencies as a function of the magnetic
moment direction. It shows that if we take the moment to be at
$\approx\!\SI{30}{\degree}$ with the $ab$ plane~\cite{Winter2016} and
small distortions towards the nearest neighbor Cl$^-$ ions both the Mu1
and Mu3 site candidates are compatible with the experimentally
observed frequencies. It should be noted that both Mu1 and Mu3 have
six nearby Cl$^-$ ions, four of which are at the $8j$ Wyckoff positions
and two of which are at the $4i$ Wyckoff positions (see
Table~\ref{tab:coulomb}).
We also considered the effect of
stacking faults at which the RuCl$_3$ layers are
translated by $\pm {\bf b}/3$~\cite{Johnson2015}.  We find that such
faults can result in a lowering of the precession frequency from muons
at the Mu1, Mu2 and Mu4 sites, but also different symmetry-equivalent
sites can become inequivalent which could be a source of broadening~\cite{SuppMat}. However, stacking faults only have a
significant effect on the precession signals if the muon is directly
adjacent to the fault~\cite{SuppMat} and so we conclude that our data are dominated
by effects due to the fault-free structure. In conclusion, the zig-zag
antiferromagnetic order with 2-layer stacking proposed by Johnson {\it
  et al.}~\cite{Johnson2015} is compatible with our \musr measurements
of \RuCl powder.  

We considered two plausible scenarios that could explain the two observed frequencies and transitions.  First, we investigated the possibility that there could be two distinct magnetic phases, one resulting from regular stacking of the layers and another from an alternative stacking proposed previously~\cite{Cao2016}.  However, our test DFT calculations~\cite{SuppMat} showed this second structure to be energetically less favorable, and moreover the second phase would not produce a distinct dipole-field signature from the first.  Second, we explored the possibility that the known presence of stacking faults~\cite{Johnson2015}, which likely lead to a complex sequence of {\it interlayer} exchange interactions, could hinder the establishing of long-range order along $k_z$.  Our simulations~\cite{SuppMat} show that a site near Mu1 is relatively insensitive to the magnetic configuration along $k_z$.  Thus, if $k_z=0.5$ order only locked in below 11\,K, a muon at this site would not be affected and would produce a precession signal all the way up to 14\,K.  However, a Mu2 or Mu3 site is found to be more sensitive to the interlayer magnetic configuration and would detect a range of frequencies if $k_z=0.5$ order is not established.  Such a site could plausibly give rise to the higher frequency signal that only sets in below 11\,K.  This second scenario is consistent with our experimental observations.

We repeated the dipole field analysis for the magnetic ordering with 3-layer stacking that Cao {\it et al.} have proposed for pristine single crystals of \RuCl~\cite{Cao2016}. While the resulting precession frequencies are all of the same order of magnitude as the experimentally observed ones, in general the 3-layer stacking leads to more than two observable frequencies to be expected, unless the frequencies due to muons stopping in the different layers and near the two types of Cl$^-$ ions are equal because of the symmetry of the muon sites~\cite{SuppMat}. We conclude that the magnetic ordering with 3-layer stacking is not compatible with our powder results, though we cannot rule out their applicability to the single crystal samples of Ref.~\cite{Cao2016}.

In conclusion, we have conducted \musr measurements of a powder of \RuCl and
confirmed a transition to long range magnetic order below $14.3(3)$~K,
with a second transition at $11.0(5)$~K. Using DFT
calculations we identified candidates for the muon stopping site and
analyzed the muon precession frequencies due to dipolar couplings at
these sites, using two zig-zag antiferromagnetic structures proposed
by recent powder and single crystal neutron diffraction studies and ab initio
calculations. After examining a number of possible scenarios, we find that our results are consistent with a 2-layer
ordering proposed by Johnson {\it et al.}~\cite{Johnson2015} and we
suggest an interpretation of our two transitions based on an
intermediate temperature regime where two-dimensional, but not
three-dimensional order, is established.

\begin{acknowledgments}
This work is supported by EPSRC (UK) grants No 1380739 and No EP/M020517/1. Y.~Li acknowledges support through a China Scholarship Council (CSC) Fellowship. R.~Valent\'{\i} thanks the Deutsche Forschungsgemeinschaft (DFG) for funding through grant SFB/TR49. We are also grateful for H.~L\"{u}tkens providing technical assistance with the experiments at PSI, A.~J.~Steele for help with the dipole field calculations, R.~Coldea for numerous insightful ideas and valuable conversations, and W.~Hayes and S.~Winter for useful discussions. 
\end{acknowledgments}
\bibliography{aRuCl3}

\begin{thebibliography}{28}%
\makeatletter
\providecommand \@ifxundefined [1]{%
 \@ifx{#1\undefined}
}%
\providecommand \@ifnum [1]{%
 \ifnum #1\expandafter \@firstoftwo
 \else \expandafter \@secondoftwo
 \fi
}%
\providecommand \@ifx [1]{%
 \ifx #1\expandafter \@firstoftwo
 \else \expandafter \@secondoftwo
 \fi
}%
\providecommand \natexlab [1]{#1}%
\providecommand \enquote  [1]{``#1''}%
\providecommand \bibnamefont  [1]{#1}%
\providecommand \bibfnamefont [1]{#1}%
\providecommand \citenamefont [1]{#1}%
\providecommand \href@noop [0]{\@secondoftwo}%
\providecommand \href [0]{\begingroup \@sanitize@url \@href}%
\providecommand \@href[1]{\@@startlink{#1}\@@href}%
\providecommand \@@href[1]{\endgroup#1\@@endlink}%
\providecommand \@sanitize@url [0]{\catcode `\\12\catcode `\$12\catcode
  `\&12\catcode `\#12\catcode `\^12\catcode `\_12\catcode `\%12\relax}%
\providecommand \@@startlink[1]{}%
\providecommand \@@endlink[0]{}%
\providecommand \url  [0]{\begingroup\@sanitize@url \@url }%
\providecommand \@url [1]{\endgroup\@href {#1}{\urlprefix }}%
\providecommand \urlprefix  [0]{URL }%
\providecommand \Eprint [0]{\href }%
\providecommand \doibase [0]{http://dx.doi.org/}%
\providecommand \selectlanguage [0]{\@gobble}%
\providecommand \bibinfo  [0]{\@secondoftwo}%
\providecommand \bibfield  [0]{\@secondoftwo}%
\providecommand \translation [1]{[#1]}%
\providecommand \BibitemOpen [0]{}%
\providecommand \bibitemStop [0]{}%
\providecommand \bibitemNoStop [0]{.\EOS\space}%
\providecommand \EOS [0]{\spacefactor3000\relax}%
\providecommand \BibitemShut  [1]{\csname bibitem#1\endcsname}%
\let\auto@bib@innerbib\@empty
\bibitem [{\citenamefont {Balents}(2010)}]{Balents2010}%
  \BibitemOpen
  \bibfield  {author} {\bibinfo {author} {\bibfnamefont {L.}~\bibnamefont
  {Balents}},\ }\href {\doibase 10.1038/nature08917} {\bibfield  {journal}
  {\bibinfo  {journal} {Nature}\ }\textbf {\bibinfo {volume} {464}},\ \bibinfo
  {pages} {199} (\bibinfo {year} {2010})}\BibitemShut {NoStop}%
\bibitem [{\citenamefont {Jackeli}\ and\ \citenamefont
  {Khaliullin}(2009)}]{Jackeli2009}%
  \BibitemOpen
  \bibfield  {author} {\bibinfo {author} {\bibfnamefont {G.}~\bibnamefont
  {Jackeli}}\ and\ \bibinfo {author} {\bibfnamefont {G.}~\bibnamefont
  {Khaliullin}},\ }\href {\doibase 10.1103/PhysRevLett.102.017205} {\bibfield
  {journal} {\bibinfo  {journal} {Physical Review Letters}\ }\textbf {\bibinfo
  {volume} {102}},\ \bibinfo {pages} {017205} (\bibinfo {year}
  {2009})}\BibitemShut {NoStop}%
\bibitem [{\citenamefont {Zhou}\ \emph {et~al.}(2016)\citenamefont {Zhou},
  \citenamefont {Li}, \citenamefont {Waugh}, \citenamefont {Parham},
  \citenamefont {Kim}, \citenamefont {Sears}, \citenamefont {Gomes},
  \citenamefont {Kee}, \citenamefont {Kim},\ and\ \citenamefont
  {Dessau}}]{Zhou2016}%
  \BibitemOpen
  \bibfield  {author} {\bibinfo {author} {\bibfnamefont {X.}~\bibnamefont
  {Zhou}}, \bibinfo {author} {\bibfnamefont {H.}~\bibnamefont {Li}}, \bibinfo
  {author} {\bibfnamefont {J.}~\bibnamefont {Waugh}}, \bibinfo {author}
  {\bibfnamefont {S.}~\bibnamefont {Parham}}, \bibinfo {author} {\bibfnamefont
  {H.-S.}\ \bibnamefont {Kim}}, \bibinfo {author} {\bibfnamefont
  {J.}~\bibnamefont {Sears}}, \bibinfo {author} {\bibfnamefont
  {A.}~\bibnamefont {Gomes}}, \bibinfo {author} {\bibfnamefont {H.-Y.}\
  \bibnamefont {Kee}}, \bibinfo {author} {\bibfnamefont {Y.-J.}\ \bibnamefont
  {Kim}}, \ and\ \bibinfo {author} {\bibfnamefont {D.}~\bibnamefont {Dessau}},\
  }\href {http://arxiv.org/abs/1603.02279} {\  (\bibinfo {year} {2016})},\
  \Eprint {http://arxiv.org/abs/1603.02279} {arXiv:1603.02279} \BibitemShut
  {NoStop}%
\bibitem [{\citenamefont {Pollini}(1994)}]{Pollini1994}%
  \BibitemOpen
  \bibfield  {author} {\bibinfo {author} {\bibfnamefont {I.}~\bibnamefont
  {Pollini}},\ }\href {\doibase 10.1103/PhysRevB.50.2095} {\bibfield  {journal}
  {\bibinfo  {journal} {Physical Review B}\ }\textbf {\bibinfo {volume} {50}},\
  \bibinfo {pages} {2095} (\bibinfo {year} {1994})}\BibitemShut {NoStop}%
\bibitem [{\citenamefont {Plumb}\ \emph {et~al.}(2014)\citenamefont {Plumb},
  \citenamefont {Clancy}, \citenamefont {Sandilands}, \citenamefont {Shankar},
  \citenamefont {Hu}, \citenamefont {Burch}, \citenamefont {Kee},\ and\
  \citenamefont {Kim}}]{Plumb2014}%
  \BibitemOpen
  \bibfield  {author} {\bibinfo {author} {\bibfnamefont {K.~W.}\ \bibnamefont
  {Plumb}}, \bibinfo {author} {\bibfnamefont {J.~P.}\ \bibnamefont {Clancy}},
  \bibinfo {author} {\bibfnamefont {L.~J.}\ \bibnamefont {Sandilands}},
  \bibinfo {author} {\bibfnamefont {V.~V.}\ \bibnamefont {Shankar}}, \bibinfo
  {author} {\bibfnamefont {Y.~F.}\ \bibnamefont {Hu}}, \bibinfo {author}
  {\bibfnamefont {K.~S.}\ \bibnamefont {Burch}}, \bibinfo {author}
  {\bibfnamefont {H.-Y.}\ \bibnamefont {Kee}}, \ and\ \bibinfo {author}
  {\bibfnamefont {Y.-J.}\ \bibnamefont {Kim}},\ }\href {\doibase
  10.1103/PhysRevB.90.041112} {\bibfield  {journal} {\bibinfo  {journal}
  {Physical Review B}\ }\textbf {\bibinfo {volume} {90}},\ \bibinfo {pages}
  {041112} (\bibinfo {year} {2014})}\BibitemShut {NoStop}%
\bibitem [{\citenamefont {Sandilands}\ \emph {et~al.}(2015)\citenamefont
  {Sandilands}, \citenamefont {Tian}, \citenamefont {Plumb}, \citenamefont
  {Kim},\ and\ \citenamefont {Burch}}]{Sandilands2015}%
  \BibitemOpen
  \bibfield  {author} {\bibinfo {author} {\bibfnamefont {L.~J.}\ \bibnamefont
  {Sandilands}}, \bibinfo {author} {\bibfnamefont {Y.}~\bibnamefont {Tian}},
  \bibinfo {author} {\bibfnamefont {K.~W.}\ \bibnamefont {Plumb}}, \bibinfo
  {author} {\bibfnamefont {Y.-J.}\ \bibnamefont {Kim}}, \ and\ \bibinfo
  {author} {\bibfnamefont {K.~S.}\ \bibnamefont {Burch}},\ }\href {\doibase
  10.1103/PhysRevLett.114.147201} {\bibfield  {journal} {\bibinfo  {journal}
  {Physical Review Letters}\ }\textbf {\bibinfo {volume} {114}},\ \bibinfo
  {pages} {147201} (\bibinfo {year} {2015})}\BibitemShut {NoStop}%
\bibitem [{\citenamefont {Banerjee}\ \emph {et~al.}(2016)\citenamefont
  {Banerjee}, \citenamefont {Bridges}, \citenamefont {Yan}, \citenamefont
  {Aczel}, \citenamefont {Li}, \citenamefont {Stone}, \citenamefont {Granroth},
  \citenamefont {Lumsden}, \citenamefont {Yiu}, \citenamefont {Knolle},
  \citenamefont {Bhattacharjee}, \citenamefont {Kovrizhin}, \citenamefont
  {Moessner}, \citenamefont {Tennant}, \citenamefont {Mandrus},\ and\
  \citenamefont {Nagler}}]{Banerjee2016}%
  \BibitemOpen
  \bibfield  {author} {\bibinfo {author} {\bibfnamefont {A.}~\bibnamefont
  {Banerjee}}, \bibinfo {author} {\bibfnamefont {C.~A.}\ \bibnamefont
  {Bridges}}, \bibinfo {author} {\bibfnamefont {J.-Q.}\ \bibnamefont {Yan}},
  \bibinfo {author} {\bibfnamefont {A.~A.}\ \bibnamefont {Aczel}}, \bibinfo
  {author} {\bibfnamefont {L.}~\bibnamefont {Li}}, \bibinfo {author}
  {\bibfnamefont {M.~B.}\ \bibnamefont {Stone}}, \bibinfo {author}
  {\bibfnamefont {G.~E.}\ \bibnamefont {Granroth}}, \bibinfo {author}
  {\bibfnamefont {M.~D.}\ \bibnamefont {Lumsden}}, \bibinfo {author}
  {\bibfnamefont {Y.}~\bibnamefont {Yiu}}, \bibinfo {author} {\bibfnamefont
  {J.}~\bibnamefont {Knolle}}, \bibinfo {author} {\bibfnamefont
  {S.}~\bibnamefont {Bhattacharjee}}, \bibinfo {author} {\bibfnamefont {D.~L.}\
  \bibnamefont {Kovrizhin}}, \bibinfo {author} {\bibfnamefont {R.}~\bibnamefont
  {Moessner}}, \bibinfo {author} {\bibfnamefont {D.~A.}\ \bibnamefont
  {Tennant}}, \bibinfo {author} {\bibfnamefont {D.~G.}\ \bibnamefont
  {Mandrus}}, \ and\ \bibinfo {author} {\bibfnamefont {S.~E.}\ \bibnamefont
  {Nagler}},\ }\href {\doibase 10.1038/nmat4604} {\bibfield  {journal}
  {\bibinfo  {journal} {Nature Materials}\ } (\bibinfo {year} {2016}),\
  10.1038/nmat4604}\BibitemShut {NoStop}%
\bibitem [{\citenamefont {Fletcher}\ \emph {et~al.}(1967)\citenamefont
  {Fletcher}, \citenamefont {Gardner}, \citenamefont {Fox},\ and\ \citenamefont
  {Topping}}]{Fletcher1967}%
  \BibitemOpen
  \bibfield  {author} {\bibinfo {author} {\bibfnamefont {J.~M.}\ \bibnamefont
  {Fletcher}}, \bibinfo {author} {\bibfnamefont {W.~E.}\ \bibnamefont
  {Gardner}}, \bibinfo {author} {\bibfnamefont {A.~C.}\ \bibnamefont {Fox}}, \
  and\ \bibinfo {author} {\bibfnamefont {G.}~\bibnamefont {Topping}},\ }\href
  {\doibase 10.1039/j19670001038} {\bibfield  {journal} {\bibinfo  {journal}
  {Journal of the Chemical Society A: Inorganic, Physical, Theoretical}\ ,\
  \bibinfo {pages} {1038}} (\bibinfo {year} {1967})}\BibitemShut {NoStop}%
\bibitem [{\citenamefont {Kobayashi}\ \emph {et~al.}(1992)\citenamefont
  {Kobayashi}, \citenamefont {Okada}, \citenamefont {Asai}, \citenamefont
  {Katada}, \citenamefont {Sano},\ and\ \citenamefont {Ambe}}]{Kobayashi1992}%
  \BibitemOpen
  \bibfield  {author} {\bibinfo {author} {\bibfnamefont {Y.}~\bibnamefont
  {Kobayashi}}, \bibinfo {author} {\bibfnamefont {T.}~\bibnamefont {Okada}},
  \bibinfo {author} {\bibfnamefont {K.}~\bibnamefont {Asai}}, \bibinfo {author}
  {\bibfnamefont {M.}~\bibnamefont {Katada}}, \bibinfo {author} {\bibfnamefont
  {H.}~\bibnamefont {Sano}}, \ and\ \bibinfo {author} {\bibfnamefont
  {F.}~\bibnamefont {Ambe}},\ }\href {\doibase 10.1021/ic00048a025} {\bibfield
  {journal} {\bibinfo  {journal} {Inorganic Chemistry}\ }\textbf {\bibinfo
  {volume} {31}},\ \bibinfo {pages} {4570} (\bibinfo {year}
  {1992})}\BibitemShut {NoStop}%
\bibitem [{\citenamefont {Fletcher}\ \emph {et~al.}(1963)\citenamefont
  {Fletcher}, \citenamefont {Gardner}, \citenamefont {Hooper}, \citenamefont
  {Hyde}, \citenamefont {Moore},\ and\ \citenamefont
  {Woodhead}}]{Fletcher1963}%
  \BibitemOpen
  \bibfield  {author} {\bibinfo {author} {\bibfnamefont {J.~M.}\ \bibnamefont
  {Fletcher}}, \bibinfo {author} {\bibfnamefont {W.~E.}\ \bibnamefont
  {Gardner}}, \bibinfo {author} {\bibfnamefont {E.~W.}\ \bibnamefont {Hooper}},
  \bibinfo {author} {\bibfnamefont {K.~R.}\ \bibnamefont {Hyde}}, \bibinfo
  {author} {\bibfnamefont {F.~H.}\ \bibnamefont {Moore}}, \ and\ \bibinfo
  {author} {\bibfnamefont {J.~L.}\ \bibnamefont {Woodhead}},\ }\href {\doibase
  10.1038/1991089a0} {\bibfield  {journal} {\bibinfo  {journal} {Nature}\
  }\textbf {\bibinfo {volume} {199}},\ \bibinfo {pages} {1089} (\bibinfo {year}
  {1963})}\BibitemShut {NoStop}%
\bibitem [{\citenamefont {Majumder}\ \emph {et~al.}(2015)\citenamefont
  {Majumder}, \citenamefont {Schmidt}, \citenamefont {Rosner}, \citenamefont
  {Tsirlin}, \citenamefont {Yasuoka},\ and\ \citenamefont
  {Baenitz}}]{Majumder2015}%
  \BibitemOpen
  \bibfield  {author} {\bibinfo {author} {\bibfnamefont {M.}~\bibnamefont
  {Majumder}}, \bibinfo {author} {\bibfnamefont {M.}~\bibnamefont {Schmidt}},
  \bibinfo {author} {\bibfnamefont {H.}~\bibnamefont {Rosner}}, \bibinfo
  {author} {\bibfnamefont {A.~A.}\ \bibnamefont {Tsirlin}}, \bibinfo {author}
  {\bibfnamefont {H.}~\bibnamefont {Yasuoka}}, \ and\ \bibinfo {author}
  {\bibfnamefont {M.}~\bibnamefont {Baenitz}},\ }\href {\doibase
  10.1103/PhysRevB.91.180401} {\bibfield  {journal} {\bibinfo  {journal}
  {Physical Review B}\ }\textbf {\bibinfo {volume} {91}},\ \bibinfo {pages}
  {180401} (\bibinfo {year} {2015})}\BibitemShut {NoStop}%
\bibitem [{\citenamefont {Sears}\ \emph {et~al.}(2015)\citenamefont {Sears},
  \citenamefont {Songvilay}, \citenamefont {Plumb}, \citenamefont {Clancy},
  \citenamefont {Qiu}, \citenamefont {Zhao}, \citenamefont {Parshall},\ and\
  \citenamefont {Kim}}]{Sears2015}%
  \BibitemOpen
  \bibfield  {author} {\bibinfo {author} {\bibfnamefont {J.~A.}\ \bibnamefont
  {Sears}}, \bibinfo {author} {\bibfnamefont {M.}~\bibnamefont {Songvilay}},
  \bibinfo {author} {\bibfnamefont {K.~W.}\ \bibnamefont {Plumb}}, \bibinfo
  {author} {\bibfnamefont {J.~P.}\ \bibnamefont {Clancy}}, \bibinfo {author}
  {\bibfnamefont {Y.}~\bibnamefont {Qiu}}, \bibinfo {author} {\bibfnamefont
  {Y.}~\bibnamefont {Zhao}}, \bibinfo {author} {\bibfnamefont {D.}~\bibnamefont
  {Parshall}}, \ and\ \bibinfo {author} {\bibfnamefont {Y.-J.}\ \bibnamefont
  {Kim}},\ }\href {\doibase 10.1103/PhysRevB.91.144420} {\bibfield  {journal}
  {\bibinfo  {journal} {Physical Review B}\ }\textbf {\bibinfo {volume} {91}},\
  \bibinfo {pages} {144420} (\bibinfo {year} {2015})}\BibitemShut {NoStop}%
\bibitem [{\citenamefont {Kubota}\ \emph {et~al.}(2015)\citenamefont {Kubota},
  \citenamefont {Tanaka}, \citenamefont {Ono}, \citenamefont {Narumi},\ and\
  \citenamefont {Kindo}}]{Kubota2015}%
  \BibitemOpen
  \bibfield  {author} {\bibinfo {author} {\bibfnamefont {Y.}~\bibnamefont
  {Kubota}}, \bibinfo {author} {\bibfnamefont {H.}~\bibnamefont {Tanaka}},
  \bibinfo {author} {\bibfnamefont {T.}~\bibnamefont {Ono}}, \bibinfo {author}
  {\bibfnamefont {Y.}~\bibnamefont {Narumi}}, \ and\ \bibinfo {author}
  {\bibfnamefont {K.}~\bibnamefont {Kindo}},\ }\href {\doibase
  10.1103/PhysRevB.91.094422} {\bibfield  {journal} {\bibinfo  {journal}
  {Physical Review B}\ }\textbf {\bibinfo {volume} {91}},\ \bibinfo {pages}
  {094422} (\bibinfo {year} {2015})}\BibitemShut {NoStop}%
\bibitem [{\citenamefont {Johnson}\ \emph {et~al.}(2015)\citenamefont
  {Johnson}, \citenamefont {Williams}, \citenamefont {Haghighirad},
  \citenamefont {Singleton}, \citenamefont {Zapf}, \citenamefont {Manuel},
  \citenamefont {Mazin}, \citenamefont {Li}, \citenamefont {Jeschke},
  \citenamefont {Valent{\'{\i}}},\ and\ \citenamefont {Coldea}}]{Johnson2015}%
  \BibitemOpen
  \bibfield  {author} {\bibinfo {author} {\bibfnamefont {R.~D.}\ \bibnamefont
  {Johnson}}, \bibinfo {author} {\bibfnamefont {S.~C.}\ \bibnamefont
  {Williams}}, \bibinfo {author} {\bibfnamefont {A.~A.}\ \bibnamefont
  {Haghighirad}}, \bibinfo {author} {\bibfnamefont {J.}~\bibnamefont
  {Singleton}}, \bibinfo {author} {\bibfnamefont {V.}~\bibnamefont {Zapf}},
  \bibinfo {author} {\bibfnamefont {P.}~\bibnamefont {Manuel}}, \bibinfo
  {author} {\bibfnamefont {I.~I.}\ \bibnamefont {Mazin}}, \bibinfo {author}
  {\bibfnamefont {Y.}~\bibnamefont {Li}}, \bibinfo {author} {\bibfnamefont
  {H.~O.}\ \bibnamefont {Jeschke}}, \bibinfo {author} {\bibfnamefont
  {R.}~\bibnamefont {Valent{\'{\i}}}}, \ and\ \bibinfo {author} {\bibfnamefont
  {R.}~\bibnamefont {Coldea}},\ }\href {\doibase 10.1103/PhysRevB.92.235119}
  {\bibfield  {journal} {\bibinfo  {journal} {Physical Review B}\ }\textbf
  {\bibinfo {volume} {92}},\ \bibinfo {pages} {235119} (\bibinfo {year}
  {2015})}\BibitemShut {NoStop}%
\bibitem [{\citenamefont {Cao}\ \emph {et~al.}(2016)\citenamefont {Cao},
  \citenamefont {Banerjee}, \citenamefont {Yan}, \citenamefont {Bridges},
  \citenamefont {Lumsden}, \citenamefont {Mandrus}, \citenamefont {Tennant},
  \citenamefont {Chakoumakos},\ and\ \citenamefont {Nagler}}]{Cao2016}%
  \BibitemOpen
  \bibfield  {author} {\bibinfo {author} {\bibfnamefont {H.~B.}\ \bibnamefont
  {Cao}}, \bibinfo {author} {\bibfnamefont {A.}~\bibnamefont {Banerjee}},
  \bibinfo {author} {\bibfnamefont {J.~Q.}\ \bibnamefont {Yan}}, \bibinfo
  {author} {\bibfnamefont {C.~A.}\ \bibnamefont {Bridges}}, \bibinfo {author}
  {\bibfnamefont {M.~D.}\ \bibnamefont {Lumsden}}, \bibinfo {author}
  {\bibfnamefont {D.~G.}\ \bibnamefont {Mandrus}}, \bibinfo {author}
  {\bibfnamefont {D.~A.}\ \bibnamefont {Tennant}}, \bibinfo {author}
  {\bibfnamefont {B.~C.}\ \bibnamefont {Chakoumakos}}, \ and\ \bibinfo {author}
  {\bibfnamefont {S.~E.}\ \bibnamefont {Nagler}},\ }\href
  {http://arxiv.org/abs/1602.08112} {\  (\bibinfo {year} {2016})},\ \Eprint
  {http://arxiv.org/abs/1602.08112} {arXiv:1602.08112} \BibitemShut {NoStop}%
\bibitem [{\citenamefont {Carretta}\ and\ \citenamefont
  {Keren}(2011)}]{Caretta2011}%
  \BibitemOpen
  \bibfield  {author} {\bibinfo {author} {\bibfnamefont {P.}~\bibnamefont
  {Carretta}}\ and\ \bibinfo {author} {\bibfnamefont {A.}~\bibnamefont
  {Keren}},\ }\href@noop {} {\emph {\bibinfo {title} {Highly Frustrated
  Magnetism}}},\ edited by\ \bibinfo {editor} {\bibfnamefont {C.}~\bibnamefont
  {Lacroix}}, \bibinfo {editor} {\bibfnamefont {P.}~\bibnamefont {Mendels}}, \
  and\ \bibinfo {editor} {\bibfnamefont {F.}~\bibnamefont {Mila}}\ (\bibinfo
  {publisher} {Springer},\ \bibinfo {address} {New York},\ \bibinfo {year}
  {2011})\ pp.\ \bibinfo {pages} {79\--106}\BibitemShut {NoStop}%
\bibitem [{Sup()}]{SuppMat}%
  \BibitemOpen
  \href@noop {} {}\bibinfo {note} {See {S}upplemental {M}aterial at [{URL} will
  be inserted by publisher] for details.}\BibitemShut {Stop}%
\bibitem [{\citenamefont {M{\"{o}}ller}\ \emph
  {et~al.}(2013{\natexlab{a}})\citenamefont {M{\"{o}}ller}, \citenamefont
  {Bonf{\`{a}}}, \citenamefont {Ceresoli}, \citenamefont {Bernardini},
  \citenamefont {Blundell}, \citenamefont {Lancaster}, \citenamefont {{De
  Renzi}}, \citenamefont {Marzari}, \citenamefont {Watanabe}, \citenamefont
  {Sulaiman},\ and\ \citenamefont {Mohamed-Ibrahim}}]{Moller2013_DFT}%
  \BibitemOpen
  \bibfield  {author} {\bibinfo {author} {\bibfnamefont {J.~S.}\ \bibnamefont
  {M{\"{o}}ller}}, \bibinfo {author} {\bibfnamefont {P.}~\bibnamefont
  {Bonf{\`{a}}}}, \bibinfo {author} {\bibfnamefont {D.}~\bibnamefont
  {Ceresoli}}, \bibinfo {author} {\bibfnamefont {F.}~\bibnamefont
  {Bernardini}}, \bibinfo {author} {\bibfnamefont {S.~J.}\ \bibnamefont
  {Blundell}}, \bibinfo {author} {\bibfnamefont {T.}~\bibnamefont {Lancaster}},
  \bibinfo {author} {\bibfnamefont {R.}~\bibnamefont {{De Renzi}}}, \bibinfo
  {author} {\bibfnamefont {N.}~\bibnamefont {Marzari}}, \bibinfo {author}
  {\bibfnamefont {I.}~\bibnamefont {Watanabe}}, \bibinfo {author}
  {\bibfnamefont {S.}~\bibnamefont {Sulaiman}}, \ and\ \bibinfo {author}
  {\bibfnamefont {M.~I.}\ \bibnamefont {Mohamed-Ibrahim}},\ }\href {\doibase
  10.1088/0031-8949/88/06/068510} {\bibfield  {journal} {\bibinfo  {journal}
  {Physica Scripta}\ ,\ \bibinfo {pages} {068510}} (\bibinfo {year}
  {2013}{\natexlab{a}})}\BibitemShut {NoStop}%
\bibitem [{\citenamefont {Foronda}\ \emph {et~al.}(2015)\citenamefont
  {Foronda}, \citenamefont {Lang}, \citenamefont {M{\"{o}}ller}, \citenamefont
  {Lancaster}, \citenamefont {Boothroyd}, \citenamefont {Pratt}, \citenamefont
  {Giblin}, \citenamefont {Prabhakaran},\ and\ \citenamefont
  {Blundell}}]{Foronda2015}%
  \BibitemOpen
  \bibfield  {author} {\bibinfo {author} {\bibfnamefont {F.~R.}\ \bibnamefont
  {Foronda}}, \bibinfo {author} {\bibfnamefont {F.}~\bibnamefont {Lang}},
  \bibinfo {author} {\bibfnamefont {J.~S.}\ \bibnamefont {M{\"{o}}ller}},
  \bibinfo {author} {\bibfnamefont {T.}~\bibnamefont {Lancaster}}, \bibinfo
  {author} {\bibfnamefont {A.~T.}\ \bibnamefont {Boothroyd}}, \bibinfo {author}
  {\bibfnamefont {F.~L.}\ \bibnamefont {Pratt}}, \bibinfo {author}
  {\bibfnamefont {S.~R.}\ \bibnamefont {Giblin}}, \bibinfo {author}
  {\bibfnamefont {D.}~\bibnamefont {Prabhakaran}}, \ and\ \bibinfo {author}
  {\bibfnamefont {S.~J.}\ \bibnamefont {Blundell}},\ }\href {\doibase
  10.1103/PhysRevLett.114.017602} {\bibfield  {journal} {\bibinfo  {journal}
  {Physical Review Letters}\ }\textbf {\bibinfo {volume} {114}},\ \bibinfo
  {pages} {017602} (\bibinfo {year} {2015})}\BibitemShut {NoStop}%
\bibitem [{\citenamefont {M{\"{o}}ller}(2013)}]{johannesthesis}%
  \BibitemOpen
  \bibfield  {author} {\bibinfo {author} {\bibfnamefont {J.}~\bibnamefont
  {M{\"{o}}ller}},\ }\emph {\bibinfo {title} {Muon-spin relaxation and its
  application in the study of molecular quantum magnets}},\ \href@noop {}
  {Ph.D. thesis},\ \bibinfo  {school} {University of Oxford} (\bibinfo {year}
  {2013})\BibitemShut {NoStop}%
\bibitem [{\citenamefont {Perdew}\ \emph {et~al.}(1996)\citenamefont {Perdew},
  \citenamefont {Burke},\ and\ \citenamefont {Ernzerhof}}]{Perdew1996}%
  \BibitemOpen
  \bibfield  {author} {\bibinfo {author} {\bibfnamefont {J.~P.}\ \bibnamefont
  {Perdew}}, \bibinfo {author} {\bibfnamefont {K.}~\bibnamefont {Burke}}, \
  and\ \bibinfo {author} {\bibfnamefont {M.}~\bibnamefont {Ernzerhof}},\ }\href
  {\doibase 10.1103/PhysRevLett.77.3865} {\bibfield  {journal} {\bibinfo
  {journal} {Physical Review Letters}\ }\textbf {\bibinfo {volume} {77}},\
  \bibinfo {pages} {3865} (\bibinfo {year} {1996})}\BibitemShut {NoStop}%
\bibitem [{\citenamefont {Blaha}\ \emph {et~al.}(2001)\citenamefont {Blaha},
  \citenamefont {Schwarz}, \citenamefont {Madsen},\ and\ \citenamefont
  {Luitz}}]{Wien2k}%
  \BibitemOpen
  \bibfield  {author} {\bibinfo {author} {\bibfnamefont {P.}~\bibnamefont
  {Blaha}}, \bibinfo {author} {\bibfnamefont {K.}~\bibnamefont {Schwarz}},
  \bibinfo {author} {\bibfnamefont {D.}~\bibnamefont {Madsen}, \bibfnamefont
  {G.~K. H.and~Kvasnicka}}, \ and\ \bibinfo {author} {\bibfnamefont
  {J.}~\bibnamefont {Luitz}},\ }\href {http://www.wien2k.at/} {\emph {\bibinfo
  {title} {{WIEN2k, \textit{An Augmented Plane Wave + Local Orbitals Program
  for Calculating Crystal Properties}}}}}\ (\bibinfo  {publisher} {Karlheinz
  Schwarz},\ \bibinfo {address} {Techn. Universit\"{a}t Wien, Austria},\
  \bibinfo {year} {2001})\BibitemShut {NoStop}%
\bibitem [{\citenamefont {Kokalj}(1999)}]{Kokalj1999}%
  \BibitemOpen
  \bibfield  {author} {\bibinfo {author} {\bibfnamefont {A.}~\bibnamefont
  {Kokalj}},\ }\href {\doibase 10.1016/S1093-3263(99)00028-5} {\bibfield
  {journal} {\bibinfo  {journal} {Journal of Molecular Graphics and Modelling}\
  }\textbf {\bibinfo {volume} {17}},\ \bibinfo {pages} {176} (\bibinfo {year}
  {1999})}\BibitemShut {NoStop}%
\bibitem [{\citenamefont {Momma}\ and\ \citenamefont
  {Izumi}(2008)}]{Momma2008}%
  \BibitemOpen
  \bibfield  {author} {\bibinfo {author} {\bibfnamefont {K.}~\bibnamefont
  {Momma}}\ and\ \bibinfo {author} {\bibfnamefont {F.}~\bibnamefont {Izumi}},\
  }\href {\doibase 10.1107/S0021889808012016} {\bibfield  {journal} {\bibinfo
  {journal} {Journal of Applied Crystallography}\ }\textbf {\bibinfo {volume}
  {41}},\ \bibinfo {pages} {653} (\bibinfo {year} {2008})}\BibitemShut
  {NoStop}%
\bibitem [{\citenamefont {M{\"{o}}ller}\ \emph
  {et~al.}(2013{\natexlab{b}})\citenamefont {M{\"{o}}ller}, \citenamefont
  {Ceresoli}, \citenamefont {Lancaster}, \citenamefont {Marzari},\ and\
  \citenamefont {Blundell}}]{Moller2013_FuF}%
  \BibitemOpen
  \bibfield  {author} {\bibinfo {author} {\bibfnamefont {J.}~\bibnamefont
  {M{\"{o}}ller}}, \bibinfo {author} {\bibfnamefont {D.}~\bibnamefont
  {Ceresoli}}, \bibinfo {author} {\bibfnamefont {T.}~\bibnamefont {Lancaster}},
  \bibinfo {author} {\bibfnamefont {N.}~\bibnamefont {Marzari}}, \ and\
  \bibinfo {author} {\bibfnamefont {S.}~\bibnamefont {Blundell}},\ }\href
  {\doibase 10.1103/PhysRevB.87.121108} {\bibfield  {journal} {\bibinfo
  {journal} {Physical Review B}\ }\textbf {\bibinfo {volume} {87}},\ \bibinfo
  {pages} {121108} (\bibinfo {year} {2013}{\natexlab{b}})}\BibitemShut
  {NoStop}%
\bibitem [{\citenamefont {Winter}\ \emph {et~al.}(2016)\citenamefont {Winter},
  \citenamefont {Li}, \citenamefont {Jeschke},\ and\ \citenamefont
  {Valenti}}]{Winter2016}%
  \BibitemOpen
  \bibfield  {author} {\bibinfo {author} {\bibfnamefont {S.~M.}\ \bibnamefont
  {Winter}}, \bibinfo {author} {\bibfnamefont {Y.}~\bibnamefont {Li}}, \bibinfo
  {author} {\bibfnamefont {H.~O.}\ \bibnamefont {Jeschke}}, \ and\ \bibinfo
  {author} {\bibfnamefont {R.}~\bibnamefont {Valenti}},\ }\href
  {http://arxiv.org/abs/1603.02548} {\  (\bibinfo {year} {2016})},\ \Eprint
  {http://arxiv.org/abs/1603.02548} {arXiv:1603.02548} \BibitemShut {NoStop}%
\bibitem [{\citenamefont {Kim}\ \emph {et~al.}(2015)\citenamefont {Kim},
  \citenamefont {V.}, \citenamefont {Catuneanu},\ and\ \citenamefont
  {Kee}}]{Kim2015}%
  \BibitemOpen
  \bibfield  {author} {\bibinfo {author} {\bibfnamefont {H.-S.}\ \bibnamefont
  {Kim}}, \bibinfo {author} {\bibfnamefont {V.~S.}\ \bibnamefont {V.}},
  \bibinfo {author} {\bibfnamefont {A.}~\bibnamefont {Catuneanu}}, \ and\
  \bibinfo {author} {\bibfnamefont {H.-Y.}\ \bibnamefont {Kee}},\ }\href
  {\doibase 10.1103/PhysRevB.91.241110} {\bibfield  {journal} {\bibinfo
  {journal} {Physical Review B}\ }\textbf {\bibinfo {volume} {91}},\ \bibinfo
  {pages} {241110} (\bibinfo {year} {2015})}\BibitemShut {NoStop}%
\bibitem [{\citenamefont {Chaloupka}\ and\ \citenamefont
  {Khaliullin}(2015)}]{Chaloupka2015}%
  \BibitemOpen
  \bibfield  {author} {\bibinfo {author} {\bibfnamefont {J.}~\bibnamefont
  {Chaloupka}}\ and\ \bibinfo {author} {\bibfnamefont {G.}~\bibnamefont
  {Khaliullin}},\ }\href {\doibase 10.1103/PhysRevB.92.024413} {\bibfield
  {journal} {\bibinfo  {journal} {Physical Review B}\ }\textbf {\bibinfo
  {volume} {92}},\ \bibinfo {pages} {024413} (\bibinfo {year}
  {2015})}\BibitemShut {NoStop}%
\end{thebibliography}%


\begin{thebibliography}{16}%
\makeatletter
\providecommand \@ifxundefined [1]{%
 \@ifx{#1\undefined}
}%
\providecommand \@ifnum [1]{%
 \ifnum #1\expandafter \@firstoftwo
 \else \expandafter \@secondoftwo
 \fi
}%
\providecommand \@ifx [1]{%
 \ifx #1\expandafter \@firstoftwo
 \else \expandafter \@secondoftwo
 \fi
}%
\providecommand \natexlab [1]{#1}%
\providecommand \enquote  [1]{``#1''}%
\providecommand \bibnamefont  [1]{#1}%
\providecommand \bibfnamefont [1]{#1}%
\providecommand \citenamefont [1]{#1}%
\providecommand \href@noop [0]{\@secondoftwo}%
\providecommand \href [0]{\begingroup \@sanitize@url \@href}%
\providecommand \@href[1]{\@@startlink{#1}\@@href}%
\providecommand \@@href[1]{\endgroup#1\@@endlink}%
\providecommand \@sanitize@url [0]{\catcode `\\12\catcode `\$12\catcode
  `\&12\catcode `\#12\catcode `\^12\catcode `\_12\catcode `\%12\relax}%
\providecommand \@@startlink[1]{}%
\providecommand \@@endlink[0]{}%
\providecommand \url  [0]{\begingroup\@sanitize@url \@url }%
\providecommand \@url [1]{\endgroup\@href {#1}{\urlprefix }}%
\providecommand \urlprefix  [0]{URL }%
\providecommand \Eprint [0]{\href }%
\providecommand \doibase [0]{http://dx.doi.org/}%
\providecommand \selectlanguage [0]{\@gobble}%
\providecommand \bibinfo  [0]{\@secondoftwo}%
\providecommand \bibfield  [0]{\@secondoftwo}%
\providecommand \translation [1]{[#1]}%
\providecommand \BibitemOpen [0]{}%
\providecommand \bibitemStop [0]{}%
\providecommand \bibitemNoStop [0]{.\EOS\space}%
\providecommand \EOS [0]{\spacefactor3000\relax}%
\providecommand \BibitemShut  [1]{\csname bibitem#1\endcsname}%
\let\auto@bib@innerbib\@empty
\bibitem [{\citenamefont {Sheldrick}(2008)}]{Sheldrick2008}%
  \BibitemOpen
  \bibfield  {author} {\bibinfo {author} {\bibfnamefont {G.~M.}\ \bibnamefont
  {Sheldrick}},\ }\href {\doibase 10.1107/S0108767307043930} {\bibfield
  {journal} {\bibinfo  {journal} {Acta Crystallographica Section A Foundations
  of Crystallography}\ }\textbf {\bibinfo {volume} {64}},\ \bibinfo {pages}
  {112} (\bibinfo {year} {2008})}\BibitemShut {NoStop}%
\bibitem [{\citenamefont {Farrugia}(2012)}]{Farrugia2012}%
  \BibitemOpen
  \bibfield  {author} {\bibinfo {author} {\bibfnamefont {L.~J.}\ \bibnamefont
  {Farrugia}},\ }\href {\doibase 10.1107/S0021889812029111} {\bibfield
  {journal} {\bibinfo  {journal} {Journal of Applied Crystallography}\ }\textbf
  {\bibinfo {volume} {45}},\ \bibinfo {pages} {849} (\bibinfo {year}
  {2012})}\BibitemShut {NoStop}%
\bibitem [{\citenamefont {Burla}\ \emph {et~al.}(2012)\citenamefont {Burla},
  \citenamefont {Caliandro}, \citenamefont {Camalli}, \citenamefont
  {Carrozzini}, \citenamefont {Cascarano}, \citenamefont {Giacovazzo},
  \citenamefont {Mallamo}, \citenamefont {Mazzone}, \citenamefont {Polidori},\
  and\ \citenamefont {Spagna}}]{Burla2012}%
  \BibitemOpen
  \bibfield  {author} {\bibinfo {author} {\bibfnamefont {M.~C.}\ \bibnamefont
  {Burla}}, \bibinfo {author} {\bibfnamefont {R.}~\bibnamefont {Caliandro}},
  \bibinfo {author} {\bibfnamefont {M.}~\bibnamefont {Camalli}}, \bibinfo
  {author} {\bibfnamefont {B.}~\bibnamefont {Carrozzini}}, \bibinfo {author}
  {\bibfnamefont {G.~L.}\ \bibnamefont {Cascarano}}, \bibinfo {author}
  {\bibfnamefont {C.}~\bibnamefont {Giacovazzo}}, \bibinfo {author}
  {\bibfnamefont {M.}~\bibnamefont {Mallamo}}, \bibinfo {author} {\bibfnamefont
  {A.}~\bibnamefont {Mazzone}}, \bibinfo {author} {\bibfnamefont
  {G.}~\bibnamefont {Polidori}}, \ and\ \bibinfo {author} {\bibfnamefont
  {R.}~\bibnamefont {Spagna}},\ }\href {\doibase 10.1107/S0021889812001124}
  {\bibfield  {journal} {\bibinfo  {journal} {Journal of Applied
  Crystallography}\ }\textbf {\bibinfo {volume} {45}},\ \bibinfo {pages} {357}
  (\bibinfo {year} {2012})}\BibitemShut {NoStop}%
\bibitem [{\citenamefont {Sheldrick}(2015)}]{Sheldrick2015}%
  \BibitemOpen
  \bibfield  {author} {\bibinfo {author} {\bibfnamefont {G.~M.}\ \bibnamefont
  {Sheldrick}},\ }\href {\doibase 10.1107/S2053229614024218} {\bibfield
  {journal} {\bibinfo  {journal} {Acta Crystallographica Section C Structural
  Chemistry}\ }\textbf {\bibinfo {volume} {71}},\ \bibinfo {pages} {3}
  (\bibinfo {year} {2015})}\BibitemShut {NoStop}%
\bibitem [{\citenamefont {Momma}\ and\ \citenamefont
  {Izumi}(2008)}]{Momma2008}%
  \BibitemOpen
  \bibfield  {author} {\bibinfo {author} {\bibfnamefont {K.}~\bibnamefont
  {Momma}}\ and\ \bibinfo {author} {\bibfnamefont {F.}~\bibnamefont {Izumi}},\
  }\href {\doibase 10.1107/S0021889808012016} {\bibfield  {journal} {\bibinfo
  {journal} {Journal of Applied Crystallography}\ }\textbf {\bibinfo {volume}
  {41}},\ \bibinfo {pages} {653} (\bibinfo {year} {2008})}\BibitemShut
  {NoStop}%
\bibitem [{\citenamefont {Johnson}\ \emph {et~al.}(2015)\citenamefont
  {Johnson}, \citenamefont {Williams}, \citenamefont {Haghighirad},
  \citenamefont {Singleton}, \citenamefont {Zapf}, \citenamefont {Manuel},
  \citenamefont {Mazin}, \citenamefont {Li}, \citenamefont {Jeschke},
  \citenamefont {Valent{\'{\i}}},\ and\ \citenamefont {Coldea}}]{Johnson2015}%
  \BibitemOpen
  \bibfield  {author} {\bibinfo {author} {\bibfnamefont {R.~D.}\ \bibnamefont
  {Johnson}}, \bibinfo {author} {\bibfnamefont {S.~C.}\ \bibnamefont
  {Williams}}, \bibinfo {author} {\bibfnamefont {A.~A.}\ \bibnamefont
  {Haghighirad}}, \bibinfo {author} {\bibfnamefont {J.}~\bibnamefont
  {Singleton}}, \bibinfo {author} {\bibfnamefont {V.}~\bibnamefont {Zapf}},
  \bibinfo {author} {\bibfnamefont {P.}~\bibnamefont {Manuel}}, \bibinfo
  {author} {\bibfnamefont {I.~I.}\ \bibnamefont {Mazin}}, \bibinfo {author}
  {\bibfnamefont {Y.}~\bibnamefont {Li}}, \bibinfo {author} {\bibfnamefont
  {H.~O.}\ \bibnamefont {Jeschke}}, \bibinfo {author} {\bibfnamefont
  {R.}~\bibnamefont {Valent{\'{\i}}}}, \ and\ \bibinfo {author} {\bibfnamefont
  {R.}~\bibnamefont {Coldea}},\ }\href {\doibase 10.1103/PhysRevB.92.235119}
  {\bibfield  {journal} {\bibinfo  {journal} {Physical Review B}\ }\textbf
  {\bibinfo {volume} {92}},\ \bibinfo {pages} {235119} (\bibinfo {year}
  {2015})}\BibitemShut {NoStop}%
\bibitem [{\citenamefont {Cao}\ \emph {et~al.}(2016)\citenamefont {Cao},
  \citenamefont {Banerjee}, \citenamefont {Yan}, \citenamefont {Bridges},
  \citenamefont {Lumsden}, \citenamefont {Mandrus}, \citenamefont {Tennant},
  \citenamefont {Chakoumakos},\ and\ \citenamefont {Nagler}}]{Cao2016}%
  \BibitemOpen
  \bibfield  {author} {\bibinfo {author} {\bibfnamefont {H.~B.}\ \bibnamefont
  {Cao}}, \bibinfo {author} {\bibfnamefont {A.}~\bibnamefont {Banerjee}},
  \bibinfo {author} {\bibfnamefont {J.~Q.}\ \bibnamefont {Yan}}, \bibinfo
  {author} {\bibfnamefont {C.~A.}\ \bibnamefont {Bridges}}, \bibinfo {author}
  {\bibfnamefont {M.~D.}\ \bibnamefont {Lumsden}}, \bibinfo {author}
  {\bibfnamefont {D.~G.}\ \bibnamefont {Mandrus}}, \bibinfo {author}
  {\bibfnamefont {D.~A.}\ \bibnamefont {Tennant}}, \bibinfo {author}
  {\bibfnamefont {B.~C.}\ \bibnamefont {Chakoumakos}}, \ and\ \bibinfo {author}
  {\bibfnamefont {S.~E.}\ \bibnamefont {Nagler}},\ }\href
  {http://arxiv.org/abs/1602.08112} {\  (\bibinfo {year} {2016})},\ \Eprint
  {http://arxiv.org/abs/1602.08112} {arXiv:1602.08112} \BibitemShut {NoStop}%
\bibitem [{\citenamefont {M{\"{o}}ller}\ \emph
  {et~al.}(2013{\natexlab{a}})\citenamefont {M{\"{o}}ller}, \citenamefont
  {Bonf{\`{a}}}, \citenamefont {Ceresoli}, \citenamefont {Bernardini},
  \citenamefont {Blundell}, \citenamefont {Lancaster}, \citenamefont {{De
  Renzi}}, \citenamefont {Marzari}, \citenamefont {Watanabe}, \citenamefont
  {Sulaiman},\ and\ \citenamefont {Mohamed-Ibrahim}}]{Moller2013_DFT}%
  \BibitemOpen
  \bibfield  {author} {\bibinfo {author} {\bibfnamefont {J.~S.}\ \bibnamefont
  {M{\"{o}}ller}}, \bibinfo {author} {\bibfnamefont {P.}~\bibnamefont
  {Bonf{\`{a}}}}, \bibinfo {author} {\bibfnamefont {D.}~\bibnamefont
  {Ceresoli}}, \bibinfo {author} {\bibfnamefont {F.}~\bibnamefont
  {Bernardini}}, \bibinfo {author} {\bibfnamefont {S.~J.}\ \bibnamefont
  {Blundell}}, \bibinfo {author} {\bibfnamefont {T.}~\bibnamefont {Lancaster}},
  \bibinfo {author} {\bibfnamefont {R.}~\bibnamefont {{De Renzi}}}, \bibinfo
  {author} {\bibfnamefont {N.}~\bibnamefont {Marzari}}, \bibinfo {author}
  {\bibfnamefont {I.}~\bibnamefont {Watanabe}}, \bibinfo {author}
  {\bibfnamefont {S.}~\bibnamefont {Sulaiman}}, \ and\ \bibinfo {author}
  {\bibfnamefont {M.~I.}\ \bibnamefont {Mohamed-Ibrahim}},\ }\href {\doibase
  10.1088/0031-8949/88/06/068510} {\bibfield  {journal} {\bibinfo  {journal}
  {Physica Scripta}\ ,\ \bibinfo {pages} {068510}} (\bibinfo {year}
  {2013}{\natexlab{a}})}\BibitemShut {NoStop}%
\bibitem [{\citenamefont {M{\"{o}}ller}\ \emph
  {et~al.}(2013{\natexlab{b}})\citenamefont {M{\"{o}}ller}, \citenamefont
  {Ceresoli}, \citenamefont {Lancaster}, \citenamefont {Marzari},\ and\
  \citenamefont {Blundell}}]{Moller2013_FuF}%
  \BibitemOpen
  \bibfield  {author} {\bibinfo {author} {\bibfnamefont {J.}~\bibnamefont
  {M{\"{o}}ller}}, \bibinfo {author} {\bibfnamefont {D.}~\bibnamefont
  {Ceresoli}}, \bibinfo {author} {\bibfnamefont {T.}~\bibnamefont {Lancaster}},
  \bibinfo {author} {\bibfnamefont {N.}~\bibnamefont {Marzari}}, \ and\
  \bibinfo {author} {\bibfnamefont {S.}~\bibnamefont {Blundell}},\ }\href
  {\doibase 10.1103/PhysRevB.87.121108} {\bibfield  {journal} {\bibinfo
  {journal} {Physical Review B}\ }\textbf {\bibinfo {volume} {87}},\ \bibinfo
  {pages} {121108} (\bibinfo {year} {2013}{\natexlab{b}})}\BibitemShut
  {NoStop}%
\bibitem [{\citenamefont {Foronda}\ \emph {et~al.}(2015)\citenamefont
  {Foronda}, \citenamefont {Lang}, \citenamefont {M{\"{o}}ller}, \citenamefont
  {Lancaster}, \citenamefont {Boothroyd}, \citenamefont {Pratt}, \citenamefont
  {Giblin}, \citenamefont {Prabhakaran},\ and\ \citenamefont
  {Blundell}}]{Foronda2015}%
  \BibitemOpen
  \bibfield  {author} {\bibinfo {author} {\bibfnamefont {F.~R.}\ \bibnamefont
  {Foronda}}, \bibinfo {author} {\bibfnamefont {F.}~\bibnamefont {Lang}},
  \bibinfo {author} {\bibfnamefont {J.~S.}\ \bibnamefont {M{\"{o}}ller}},
  \bibinfo {author} {\bibfnamefont {T.}~\bibnamefont {Lancaster}}, \bibinfo
  {author} {\bibfnamefont {A.~T.}\ \bibnamefont {Boothroyd}}, \bibinfo {author}
  {\bibfnamefont {F.~L.}\ \bibnamefont {Pratt}}, \bibinfo {author}
  {\bibfnamefont {S.~R.}\ \bibnamefont {Giblin}}, \bibinfo {author}
  {\bibfnamefont {D.}~\bibnamefont {Prabhakaran}}, \ and\ \bibinfo {author}
  {\bibfnamefont {S.~J.}\ \bibnamefont {Blundell}},\ }\href {\doibase
  10.1103/PhysRevLett.114.017602} {\bibfield  {journal} {\bibinfo  {journal}
  {Physical Review Letters}\ }\textbf {\bibinfo {volume} {114}},\ \bibinfo
  {pages} {017602} (\bibinfo {year} {2015})}\BibitemShut {NoStop}%
\bibitem [{\citenamefont {Winter}\ \emph {et~al.}(2016)\citenamefont {Winter},
  \citenamefont {Li}, \citenamefont {Jeschke},\ and\ \citenamefont
  {Valenti}}]{Winter2016}%
  \BibitemOpen
  \bibfield  {author} {\bibinfo {author} {\bibfnamefont {S.~M.}\ \bibnamefont
  {Winter}}, \bibinfo {author} {\bibfnamefont {Y.}~\bibnamefont {Li}}, \bibinfo
  {author} {\bibfnamefont {H.~O.}\ \bibnamefont {Jeschke}}, \ and\ \bibinfo
  {author} {\bibfnamefont {R.}~\bibnamefont {Valenti}},\ }\href
  {http://arxiv.org/abs/1603.02548} {\  (\bibinfo {year} {2016})},\ \Eprint
  {http://arxiv.org/abs/1603.02548} {arXiv:1603.02548} \BibitemShut {NoStop}%
\bibitem [{\citenamefont {Banerjee}\ \emph {et~al.}(2016)\citenamefont
  {Banerjee}, \citenamefont {Bridges}, \citenamefont {Yan}, \citenamefont
  {Aczel}, \citenamefont {Li}, \citenamefont {Stone}, \citenamefont {Granroth},
  \citenamefont {Lumsden}, \citenamefont {Yiu}, \citenamefont {Knolle},
  \citenamefont {Bhattacharjee}, \citenamefont {Kovrizhin}, \citenamefont
  {Moessner}, \citenamefont {Tennant}, \citenamefont {Mandrus},\ and\
  \citenamefont {Nagler}}]{Banerjee2016}%
  \BibitemOpen
  \bibfield  {author} {\bibinfo {author} {\bibfnamefont {A.}~\bibnamefont
  {Banerjee}}, \bibinfo {author} {\bibfnamefont {C.~A.}\ \bibnamefont
  {Bridges}}, \bibinfo {author} {\bibfnamefont {J.-Q.}\ \bibnamefont {Yan}},
  \bibinfo {author} {\bibfnamefont {A.~A.}\ \bibnamefont {Aczel}}, \bibinfo
  {author} {\bibfnamefont {L.}~\bibnamefont {Li}}, \bibinfo {author}
  {\bibfnamefont {M.~B.}\ \bibnamefont {Stone}}, \bibinfo {author}
  {\bibfnamefont {G.~E.}\ \bibnamefont {Granroth}}, \bibinfo {author}
  {\bibfnamefont {M.~D.}\ \bibnamefont {Lumsden}}, \bibinfo {author}
  {\bibfnamefont {Y.}~\bibnamefont {Yiu}}, \bibinfo {author} {\bibfnamefont
  {J.}~\bibnamefont {Knolle}}, \bibinfo {author} {\bibfnamefont
  {S.}~\bibnamefont {Bhattacharjee}}, \bibinfo {author} {\bibfnamefont {D.~L.}\
  \bibnamefont {Kovrizhin}}, \bibinfo {author} {\bibfnamefont {R.}~\bibnamefont
  {Moessner}}, \bibinfo {author} {\bibfnamefont {D.~A.}\ \bibnamefont
  {Tennant}}, \bibinfo {author} {\bibfnamefont {D.~G.}\ \bibnamefont
  {Mandrus}}, \ and\ \bibinfo {author} {\bibfnamefont {S.~E.}\ \bibnamefont
  {Nagler}},\ }\href {\doibase 10.1038/nmat4604} {\bibfield  {journal}
  {\bibinfo  {journal} {Nature Materials}\ } (\bibinfo {year} {2016}),\
  10.1038/nmat4604}\BibitemShut {NoStop}%
\bibitem [{\citenamefont {Blaha}\ \emph {et~al.}(2001)\citenamefont {Blaha},
  \citenamefont {Schwarz}, \citenamefont {Madsen},\ and\ \citenamefont
  {Luitz}}]{Wien2k}%
  \BibitemOpen
  \bibfield  {author} {\bibinfo {author} {\bibfnamefont {P.}~\bibnamefont
  {Blaha}}, \bibinfo {author} {\bibfnamefont {K.}~\bibnamefont {Schwarz}},
  \bibinfo {author} {\bibfnamefont {D.}~\bibnamefont {Madsen}, \bibfnamefont
  {G.~K. H.and~Kvasnicka}}, \ and\ \bibinfo {author} {\bibfnamefont
  {J.}~\bibnamefont {Luitz}},\ }\href {http://www.wien2k.at/} {\emph {\bibinfo
  {title} {{WIEN2k, \textit{An Augmented Plane Wave + Local Orbitals Program
  for Calculating Crystal Properties}}}}}\ (\bibinfo  {publisher} {Karlheinz
  Schwarz},\ \bibinfo {address} {Techn. Universit\"{a}t Wien, Austria},\
  \bibinfo {year} {2001})\BibitemShut {NoStop}%
\bibitem [{\citenamefont {Koepernik}\ and\ \citenamefont
  {Eschrig}(1999)}]{FPLO1}%
  \BibitemOpen
  \bibfield  {author} {\bibinfo {author} {\bibfnamefont {K.}~\bibnamefont
  {Koepernik}}\ and\ \bibinfo {author} {\bibfnamefont {H.}~\bibnamefont
  {Eschrig}},\ }\href {\doibase 10.1103/PhysRevB.59.1743} {\bibfield  {journal}
  {\bibinfo  {journal} {Phys. Rev. B}\ }\textbf {\bibinfo {volume} {59}},\
  \bibinfo {pages} {1743} (\bibinfo {year} {1999})}\BibitemShut {NoStop}%
\bibitem [{\citenamefont {Opahle}\ \emph {et~al.}(1999)\citenamefont {Opahle},
  \citenamefont {Koepernik},\ and\ \citenamefont {Eschrig}}]{FPLO2}%
  \BibitemOpen
  \bibfield  {author} {\bibinfo {author} {\bibfnamefont {I.}~\bibnamefont
  {Opahle}}, \bibinfo {author} {\bibfnamefont {K.}~\bibnamefont {Koepernik}}, \
  and\ \bibinfo {author} {\bibfnamefont {H.}~\bibnamefont {Eschrig}},\ }\href
  {\doibase 10.1103/PhysRevB.60.14035} {\bibfield  {journal} {\bibinfo
  {journal} {Phys. Rev. B}\ }\textbf {\bibinfo {volume} {60}},\ \bibinfo
  {pages} {14035} (\bibinfo {year} {1999})}\BibitemShut {NoStop}%
\bibitem [{\citenamefont {Pollini}(1994)}]{Pollini1994}%
  \BibitemOpen
  \bibfield  {author} {\bibinfo {author} {\bibfnamefont {I.}~\bibnamefont
  {Pollini}},\ }\href {\doibase 10.1103/PhysRevB.50.2095} {\bibfield  {journal}
  {\bibinfo  {journal} {Physical Review B}\ }\textbf {\bibinfo {volume} {50}},\
  \bibinfo {pages} {2095} (\bibinfo {year} {1994})}\BibitemShut {NoStop}%
\end{thebibliography}%
\end{document}



\title{Novel magnetism on a honeycomb lattice in \RuCl studied by muon spin rotation
\newline
{\it Supplemental Material}}

\author{F.~Lang}
\email{franz.lang@physics.ox.ac.uk}
\affiliation{Oxford University Department of Physics, Clarendon Laboratory, Parks Road, Oxford, OX1 3PU, United Kingdom}

\author{ P.~J.~Baker}
\affiliation{ISIS Facility, Rutherford Appleton Laboratory, Chilton, Oxfordshire OX11 0QX, United Kingdom}

\author{ A.~A.~Haghighirad}
\affiliation{Oxford University Department of Physics, Clarendon Laboratory, Parks Road, Oxford, OX1 3PU, United Kingdom}

\author{Y.~Li}
\affiliation{Institut f\"{u}r Theoretische Physik, Goethe-Universit\"{a}t Frankfurt, 60438 Frankfurt am Main, Germany}

\author{D.~Prabhakaran}
\affiliation{Oxford University Department of Physics, Clarendon Laboratory, Parks Road, Oxford, OX1 3PU, United Kingdom}

\author{R.~Valent\'{\i}}
\affiliation{Institut f\"{u}r Theoretische Physik, Goethe-Universit\"{a}t Frankfurt, 60438 Frankfurt am Main, Germany}

\author{S.~J.~Blundell}
\email{s.blundell@physics.ox.ac.uk}
\affiliation{Oxford University Department of Physics, Clarendon Laboratory, Parks Road, Oxford, OX1 3PU, United Kingdom}

\date{\today}

\begin{abstract}
Here additional information is provided for \ref{sec:xstalstruct}. the characterization of our \RuCl sample using x-ray diffraction, \ref{sec:bond}. an estimate for the length of the Cl--$\mu$ bond expected in \RuCl, and \ref{sec:dipoles}. an analysis of the impact of stacking faults and different proposed  magnetic and crystal structures on the dipolar fields at the muon site candidates.
\end{abstract}

\maketitle

\section{Crystal structure characterization\label{sec:xstalstruct}}
\begin{figure}[htbp] \center
\includegraphics[width=\columnwidth, clip, trim= 0.0mm 0.0mm 1.0mm 0.0mm]{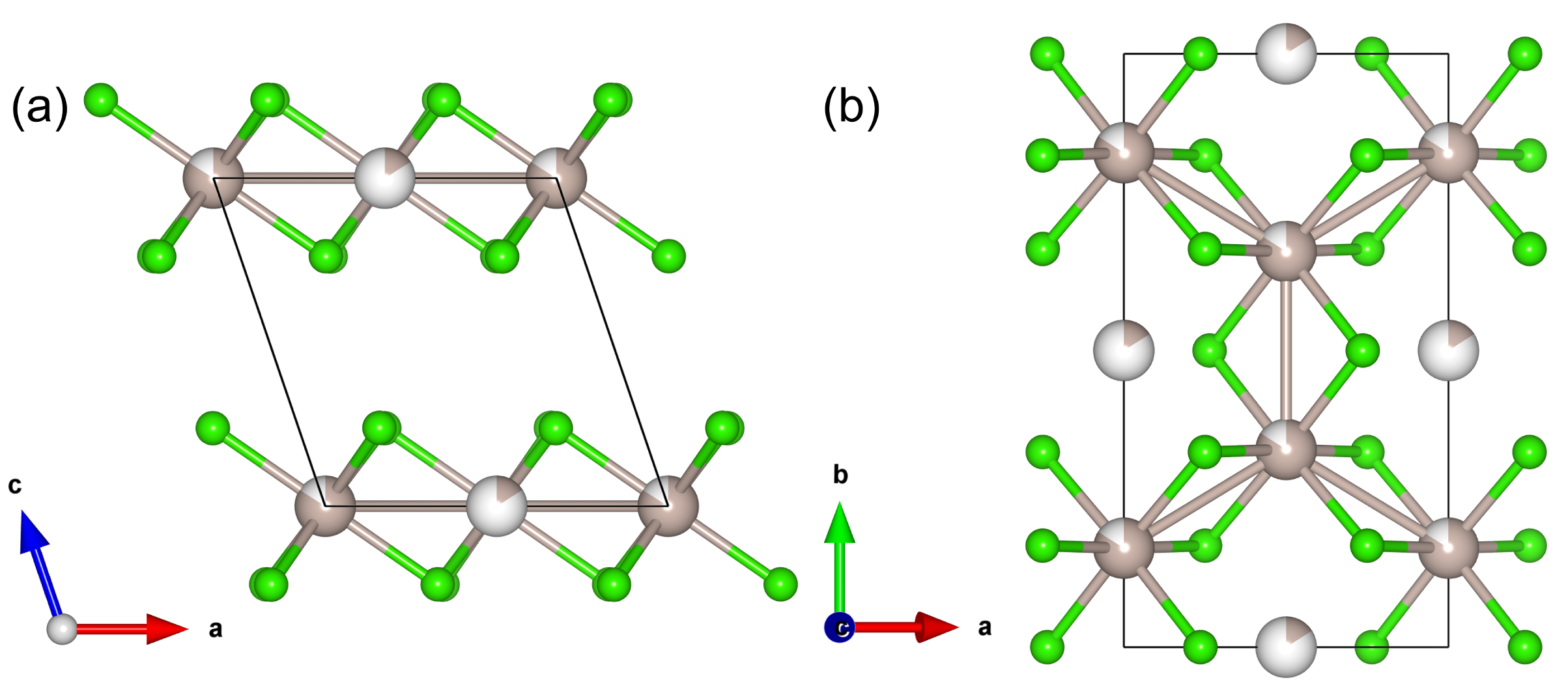}
\caption{ \label{plot:xstalstruct}
\RuCl crystal structure, showing Ru atoms as grey spheres, Cl as green spheres and the unit cell as a black outline. The partial coloring of the grey spheres indicates the occupational fraction of the corresponding Ru site (see Table~\ref{tab:xstaldata}), with which the stacking faults by $\pm {\bf b}/3$ of the Ru layers was modeled. Panels (a) and (b) show projections along the $b$ and $c$ axes, respectively.
}
\end{figure}

\begin{figure}[htbp] \center
\includegraphics[width=\columnwidth, clip, trim= 0.0mm 0.0mm 0.0mm 0.0mm]{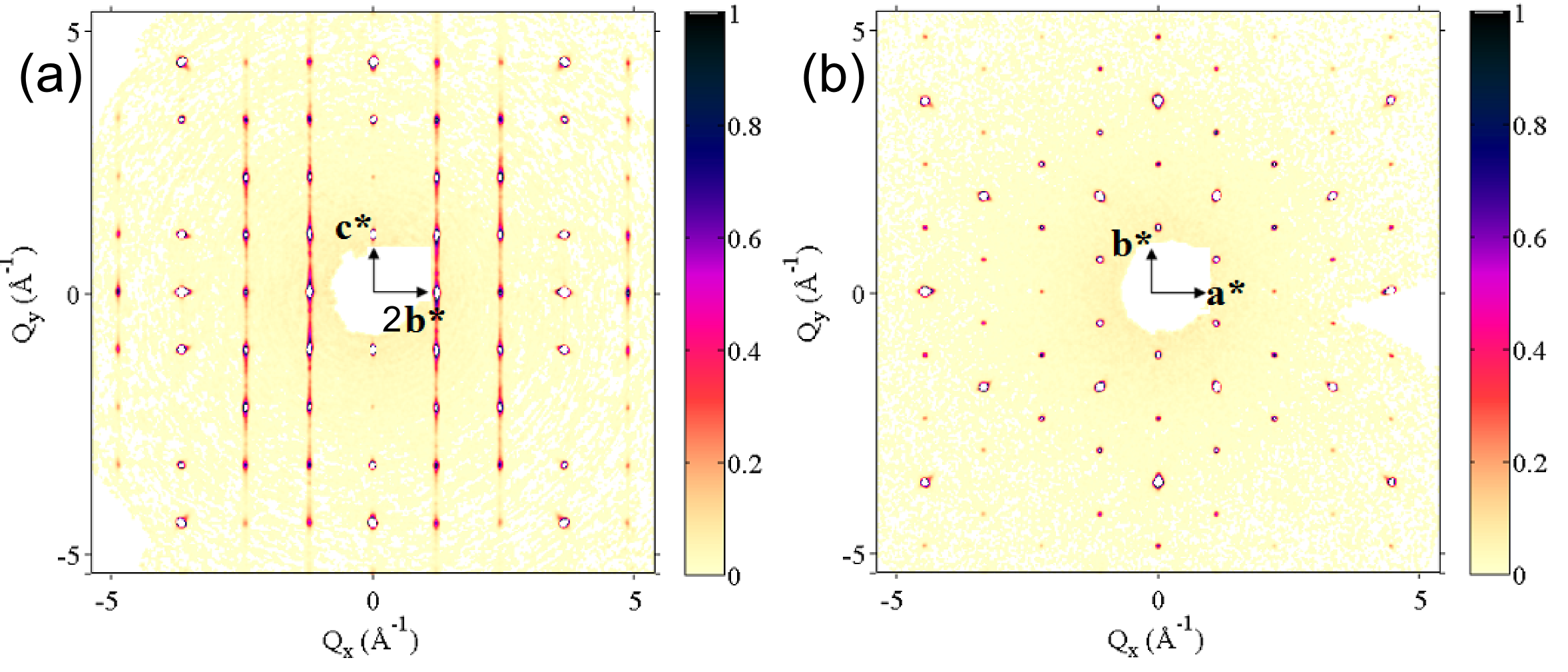}
\caption{ \label{plot:diffpattern}
X-ray diffraction patterns showing the $0kl$ and $hk0$ planes in panels (a) and (b), respectively.
}
\end{figure}

Single crystal X-ray diffraction data reported here were collected with molybdenum K--$\alpha$ radiation ($\lambda = \SI{0.71073}{\angstrom}$) using an Agilent Supernova diffractometer equipped with an Atlas detector. We performed the data integration and cell refinement using the CrysAlis Pro Software, analyzed the structure by SIR-2011 in WinGX, and refined the data using the SHELXL 2014 software package~\cite{Sheldrick2008,Farrugia2012,Burla2012,Sheldrick2015}. Visualisation of the crystal structure (Figure~\ref{plot:xstalstruct}) was done in VESTA~\cite{Momma2008}.  
The observed diffraction pattern of a small single crystal, picked out of the polycrystalline batch used for our \musr measurements, is fully consistent with the structural model of Ref.~\cite{Johnson2015}, whereby honeycomb layers of edge-sharing RuCl$_6$ octahedra are vertically stacked with an offset along $-{\bf a}$ in a monoclinic crystal structure (spacegroup $C2/m$). In the diffraction patterns, presented in Figure~\ref{plot:diffpattern}, diffuse scattering is visible in addition to sharp Bragg-reflections along $l$ with general selection rule $k = 3n + 1$ and $3n + 2$ ($n$ is integer) and $h + k = 2n$ (due to $C$-centering). This diffuse scattering has been attributed to the occurrence of stacking faults~\cite{Johnson2015}, whereby a layer is in-plane shifted by $\pm {\bf b}/3$. 
The associated energy cost is minimal as the Cl positions are unchanged upon such a shift. 
In the structural refinement of the sharp diffraction peaks we parameterize such $\pm {\bf b}/3$ stacking faults by allowing Ru atoms to partially occupy the honeycomb center site $(0.5,0,0)$ at the expense of the nominal Ru site $(0.5,0.33325,0)$.The results of the converged structural refinement at room temperature are shown in Table~\ref{tab:xstaldata}. The fractional site occupancies for the two Ru positions show that we roughly have about one stacking fault every six Ru layers.

\begin{table*}[htb]
\begin{tabularx}{\textwidth}{@{\extracolsep{\fill}}*2 l@{\extracolsep{\fill}}*5c }
Compound &\multicolumn{6}{l}{\RuCl} \\
\hline\\[-10pt]
Measurement temperature &\multicolumn{6}{l}{$293$~K} \\
\hline\\[-10pt]
Crystal system & \multicolumn{6}{l}{Monoclinic} \\
\hline\\[-10pt]
Space group & \multicolumn{6}{l}{$C2/m$} \\
\hline\\[-10pt]
Unit cell dimensions	& \multicolumn{6}{l}{$a=5.985\pm\SI{0.005}{\angstrom}$, $b=10.355\pm\SI{0.005}{\angstrom}$, $c=6.049\pm\SI{0.005}{\angstrom}$,}\\
 & \multicolumn{6}{l}{$\alpha=\gamma=\SI{90}{\degree}$, $\beta=\SI{108.830}{\degree}$, Volumme$=\SI{354.82}{\angstrom}^3$} \\
\hline\\[-10pt]
Z & \multicolumn{6}{l}{4}\\
\hline\\[-10pt]
Density (calculated) & \multicolumn{6}{l}{$3.883$~g/cm$^3$}\\
\hline\\[-10pt]
Reflections collected & \multicolumn{6}{l}{$5431$}\\
\hline\\[-10pt]
Unique reflections & \multicolumn{6}{l}{ $312$ of which $0$ suppressed} \\
\hline\\[-10pt]
R(int) & \multicolumn{6}{l}{$0.0481$} \\
\hline\\[-10pt]
R(sigma) & \multicolumn{6}{l}{$0.0184$} \\
\hline\\[-10pt]
Goodness-of-fit & \multicolumn{6}{l}{$1.192$} \\
\hline\\[-10pt]
Final R indices (R$_\text{all}$) & \multicolumn{6}{l}{$0.0559$} \\
\hline\\[-10pt]
wR$_\text{obs}$ & \multicolumn{6}{l}{$0.1359$} \\
\hline\\[-10pt]
Wavelength & \multicolumn{6}{l}{$\SI{0.71073}{\angstrom}$} \\
\hline\\[-10pt]
Weight scheme for & \multicolumn{6}{l}{$\text{Weight}=1/ [\text{sigma}^2(\text{Fo}^2) + ( 0.0964 * \text{P})^2 + 1.48 * \text{P} ]$,}\\
 the refinement &\multicolumn{6}{l}{where $\text{P}=(\text{Max}(\text{Fo}^2,0)+2* \text{Fc}^2)/3$}\\
\hline\\[-10pt]
Atomic Wyckoff-positions & Atom & Site & x & y & z & site occupancy\\
&Ru &           4j   &      0.50  &    0.33325   &      0       &          0.864\\
&Ru   &         2b   &     0.50      &       0       &       0         &        0.13578  \\
&Cl &            8j     &    0.2501 &  0.17107  &   0.76279   &     1\\
&Cl   &          4i    &     0.2359    &     0       &    0.23865    &     1\\
\hline\\[-10pt]
Isotropic temperature factors ($\SI{}{\angstrom}^2$) & \multicolumn{6}{l}{$U_{iso}(\text{Ru}) 0.01678\pm0.00064, (\text{Ru}) 0.05323\pm0.00189$,} \\
& \multicolumn{6}{l}{$(\text{Cl}) 0.01128\pm0.00068, (\text{Cl})0.01176\pm0.00071$}\\
\hline\\[-10pt]
Anisotropic temperature factor ($\SI{}{\angstrom}^2$) & \multicolumn{6}{l}{$U_{11}(\text{Ru})=0.01750\pm0.00095, U_{11}(\text{Ru})=0.04896\pm0.00379$,}\\
& \multicolumn{6}{l}{$U_{11}(\text{Cl})=0.01162\pm0.00112, U_{11}(\text{Cl}) = 0.01466 \pm 0.00133$,} \\
& \multicolumn{6}{l}{$U_{22}(\text{Ru}) = 0.01431 \pm 0.00096, U_{22}(\text{Ru}) = 0.05372 \pm 0.00422$,} \\
& \multicolumn{6}{l}{$U_{22}(\text{Cl}) = 0.01316 \pm 0.00116, U_{22}(\text{Cl}) = 0.00920 \pm 0.00127$,} \\
& \multicolumn{6}{l}{$U_{33}(\text{Ru}) = 0.01916 \pm 0.00097, U_{33}(\text{Ru}) = 0.05909 \pm 0.00427$, }\\
&\multicolumn{6}{l}{ $U_{33}(\text{Cl}) = 0.00878 \pm 0.00120, U_{33}(\text{Cl}) = 0.00934 \pm 0.00134$,} \\
& \multicolumn{6}{l}{$U_{13}(\text{Ru}) = 0.00680 \pm 0.00062, U_{13}(\text{Ru}) = 0.02031 \pm 0.00311$,}\\
&\multicolumn{6}{l}{ $U_{13}(\text{Cl}) = 0.00291 \pm 0.00081, U_{13}(\text{Cl}) = 0.00097 \pm 0.00095$}\\
\hline\\[-10pt]
\end{tabularx}
\caption{ \label{tab:xstaldata}
Structural parameters of \RuCl at room temperature by single crystal X-ray diffraction. Note that the fractional coordinates are in agreement with previously published ones~\cite{Johnson2015,Cao2016} upon a change of origin by $(0.5,0,0)$.
}
\end{table*}

\section{Cl--$\mu$ bond length estimate\label{sec:bond}}
 Previous \musr measurements involving compounds containing fluorine and oxygen~\cite{Moller2013_DFT} lead us to expect the formation of a Cl--$\mu$ bond in \RuCl with a bond length that can be estimated as follows. Using the known lengths for H--F ($\SI{0.92}{\angstrom}$), H--Cl ($\SI{1.27}{\angstrom}$) and H--O ($\SI{0.96}{\angstrom}$) bonds and simply scaling, using the bond lengths of F--$\mu$ ($1.14$--$\SI{1.21}{\angstrom}$)~\cite{Moller2013_FuF} and O--$\mu$ ($\approx\!\SI{1.0}{\angstrom}$)~\cite{Moller2013_FuF, Foronda2015}, we obtain a likely Cl--$\mu$ bond length of roughly $1.5$--$\SI{1.6}{\angstrom}$.

\section{Dipole field calculations\label{sec:dipoles}}
We note that
both Mu1 and Mu3 sites which we identified have
six nearby Cl$^-$ ions, four of which are at the $8j$ Wyckoff positions
and two of which are at the $4i$ Wyckoff positions.  Consequently, if
either Mu1 or Mu3 is indeed close to the correct muon stopping site we
would expect two muon precession frequencies with an amplitude ratio of 2 : 1.
This is roughly in line with what is observed with the two
frequencies, but this would not explain why one frequency disappears
at around 11\,K.  Consequently, we have explored some other possible
scenarios.
\subsection{Stacking faults}
To study the effect of stacking faults at which the RuCl$_3$ layers
are translated by $\pm {\bf b}/3$ (see Ref.~\onlinecite{Johnson2015}
for details), we calculated the dipolar field at our candidate muon
sites assuming the presence of stacking faults. For these
calculations, we chose the magnetic structure of Ref.~\cite{Johnson2015}
and assumed, following Ref.~\cite{Winter2016}, that the moments lie at
$\SI{30}{\degree}$ to the $a$-axis in the $ac$ plane.  We calculated
our results assuming that the muon site was somewhere between the
sites identified by our electrostatic calculations and the nearest
Cl$^-$ ion, parametrizing these sites by the distance along this line
from $0$--$\SI{1}{\angstrom}$ (we expect the likely site to be
somewhere between these limits).  In the presence of stacking faults,
the symmetry between certain sites is broken, so that we needed to
calculate the dipolar field for all the 8j and 4i sites close to our
candidate sites. Our results are shown in
Figure~\ref{plot:dipolestacking}. The first column of panels shows the
results without stacking faults, while subsequent columns show results
with stacking faults at the $z$-values indicated (note that the Ru
planes lie at integer values of $z$, and so we notionally indicate the
stacking faults at half-integer values). These results demonstrate
that stacking faults only have a significant effect on the precession
signals if the muon is directly adjacent to the fault. For example
Mu1, which is at $z=0$, is affected by stacking faults at
$z=\pm\frac{1}{2}$, whereas Mu3 and Mu4, which are at $z=\frac{1}{2}$,
are primarily affected by a stacking fault at $z=\frac{1}{2}$. Thus,
the effect of stacking faults on the muon precession signal is very
local. Notice also that different symmetry-equivalent sites (which
result in identical traces for the case without stacking faults) can
become inequivalent when a stacking fault is nearby and this could be
a source of broadening (because the slightly different frequencies
will sum together to produce a damped signal). 

\subsection{Magnetization reversal at stacking faults}
It is also possible that at a stacking fault the magnetic moments
reverse in sign (this cannot be determined from the data in
Ref.~\onlinecite{Johnson2015}, but this scenario is possible given the change in
exchange pathways at the fault).  We have considered this possibility
and repeated the above calculations for the case in which all magnetic
moments are reversed (${\bf m}\to -{\bf m}$) below the
stacking fault.  The corresponding calculations are shown in
Figure~\ref{plot:dipolestacking2}.  The results are quite similar, but
there are noticeable differences when the fault is close to the muon,
and in particular in this case Mu3 sites have lower precession signals
while those at Mu2 are higher.

\begin{figure*}[htbp] \center
\includegraphics[width=\textwidth, clip, trim= 4.0mm 10.0mm 5.0mm 7.0mm]{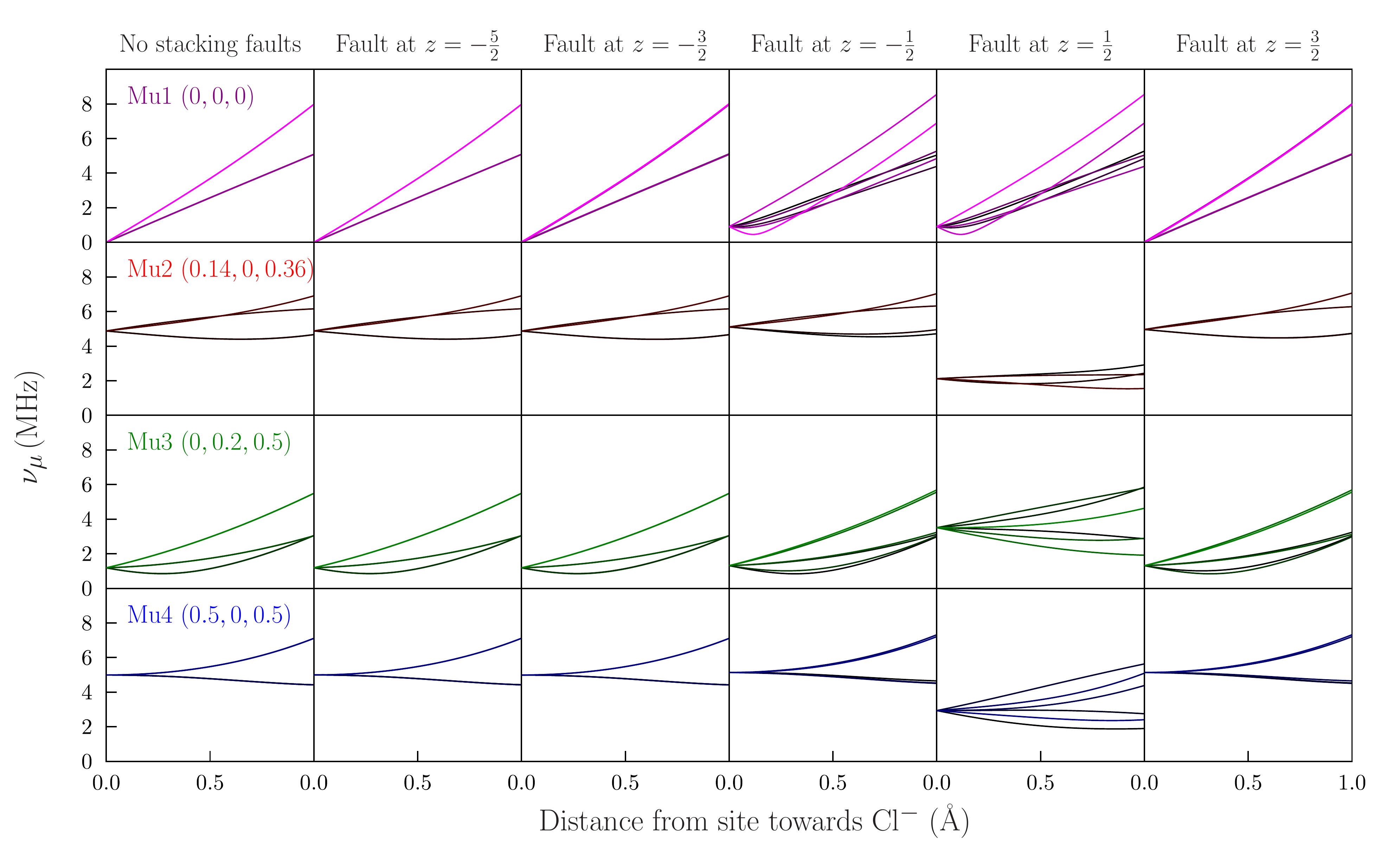}
\caption{ \label{plot:dipolestacking}
Muon precession frequencies for muon site candidates near a stacking fault. See the main text for details.
}
\end{figure*}

\begin{figure*}[htbp] \center
\includegraphics[width=\textwidth, clip, trim= 4.0mm 10.0mm 5.0mm 7.0mm]{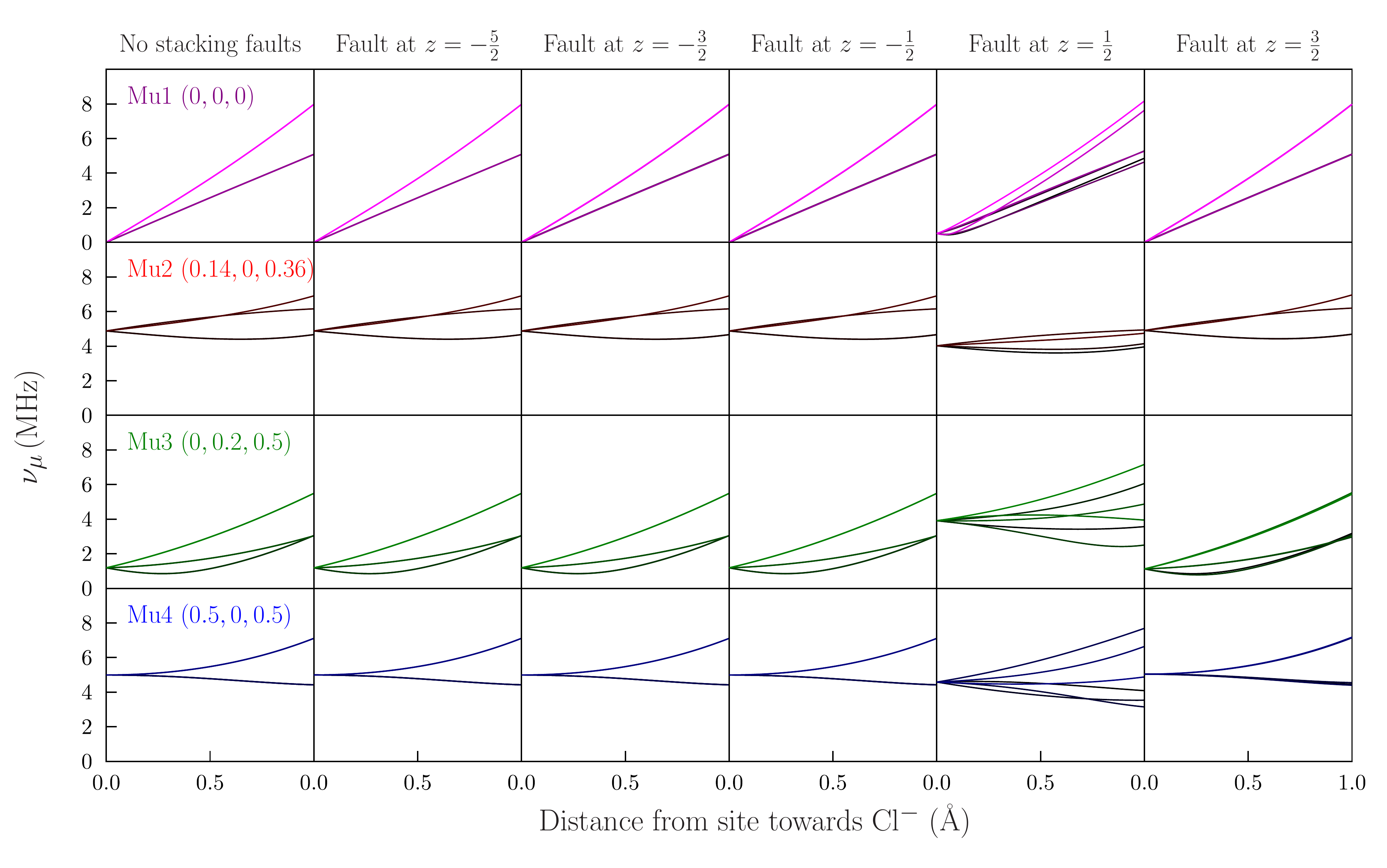}
\caption{ \label{plot:dipolestacking2}
Muon precession frequencies as for Figure~\ref{plot:dipolestacking},
but with spins reversed at a stacking fault.
}
\end{figure*}

However, since the stacking faults only have a significant effect on
the precession signals if the muon is very close by (in either case
illustrated in Figure~\ref{plot:dipolestacking} or
Figure~\ref{plot:dipolestacking2}), and since the distance between
stacking faults may be of the order of 5--6 lattice planes
(see~\cite{Johnson2015} or Section~\ref{sec:xstalstruct} above), we
conclude that our data are dominated by effects due to the fault-free
structure.

\subsection{In-plane order and three-dimensional order}
\begin{figure*}[htbp] \center
\includegraphics[width=\textwidth, clip, trim= 4.0mm 10.0mm 5.0mm 7.0mm]{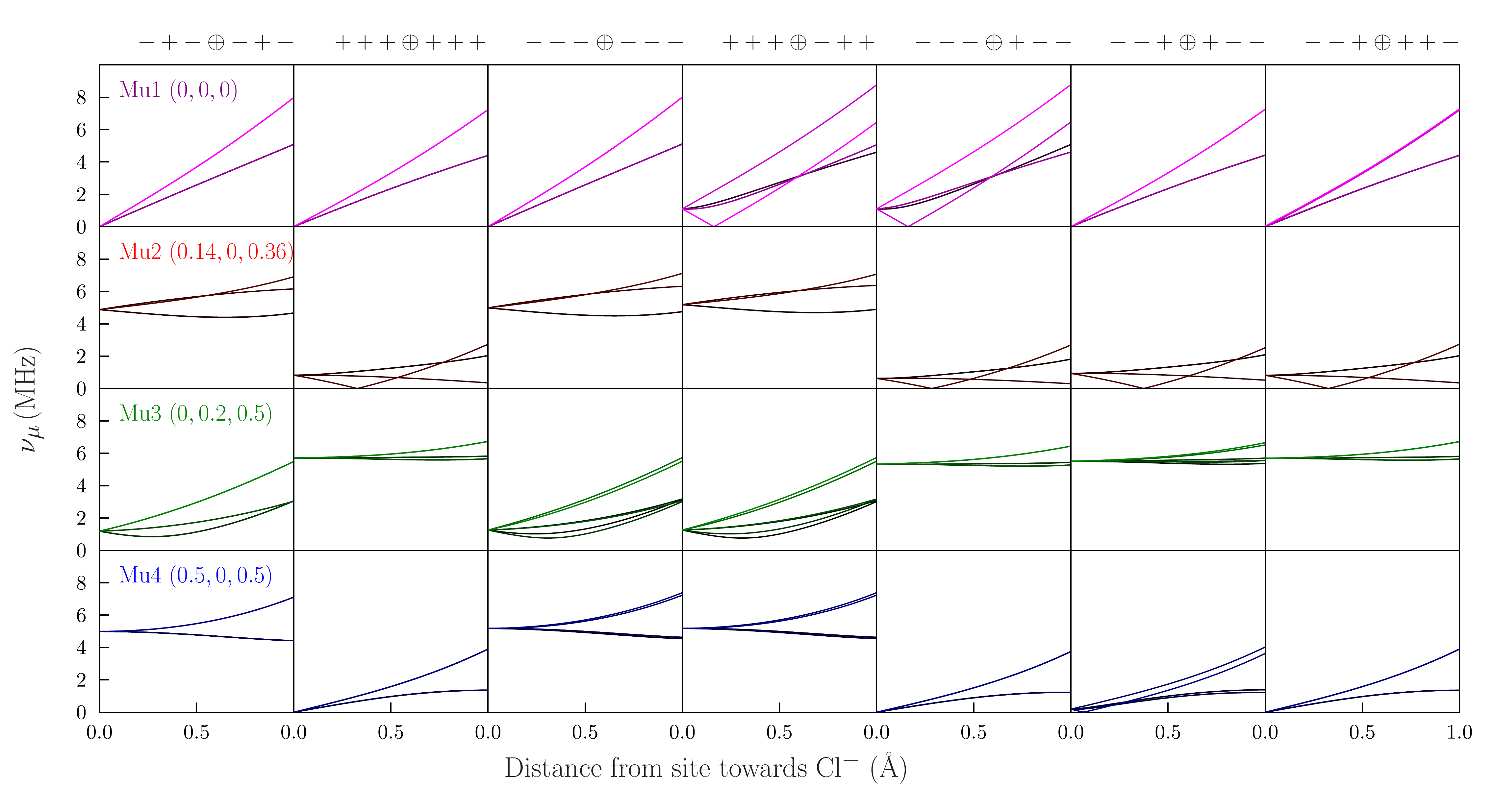}
\caption{ \label{plot:random}
Muon precession frequencies as for Figure~\ref{plot:dipolestacking},
but now with different $c$-axis ordering configurations.  These are
indicated schematically by the sequence above each column, with the
muon in the layer indicated by $\oplus$ and the nearby layers
indicated with $+$ and $-$.  The low-temperature magnetic structure
proposed in Ref.~\onlinecite{Johnson2015}
corresponds to the column on the left ($-+-\oplus -+-$).
}
\end{figure*}

In the magnetic structure proposed in Ref.~\onlinecite{Johnson2015}
the spins in neighboring layers lie antiparallel.  To examine the
effect of in-plane order, but lack of order between the planes, we
calculate the dipole fields at the candidate muon sites with different
ordering configurations along the $c$-axis.
These are shown in
Fig.~\ref{plot:random}.  These results demonstrate that the
precession signals at Mu1 are largely unaffected by lack of $c$-axis
order, though the greatest effect is when the magnetic configuration
around the muon site is asymmetric with respect to a reflection about
the plane $z=0$.  The precession signals corresponding to the other
sites are more strongly affected.


\subsection{AB stacking of Ru layers}
\begin{figure*}[htbp] \center
\includegraphics[width=1.1\columnwidth, clip, trim= 0.0mm 10.0mm 5.0mm 2.0mm]{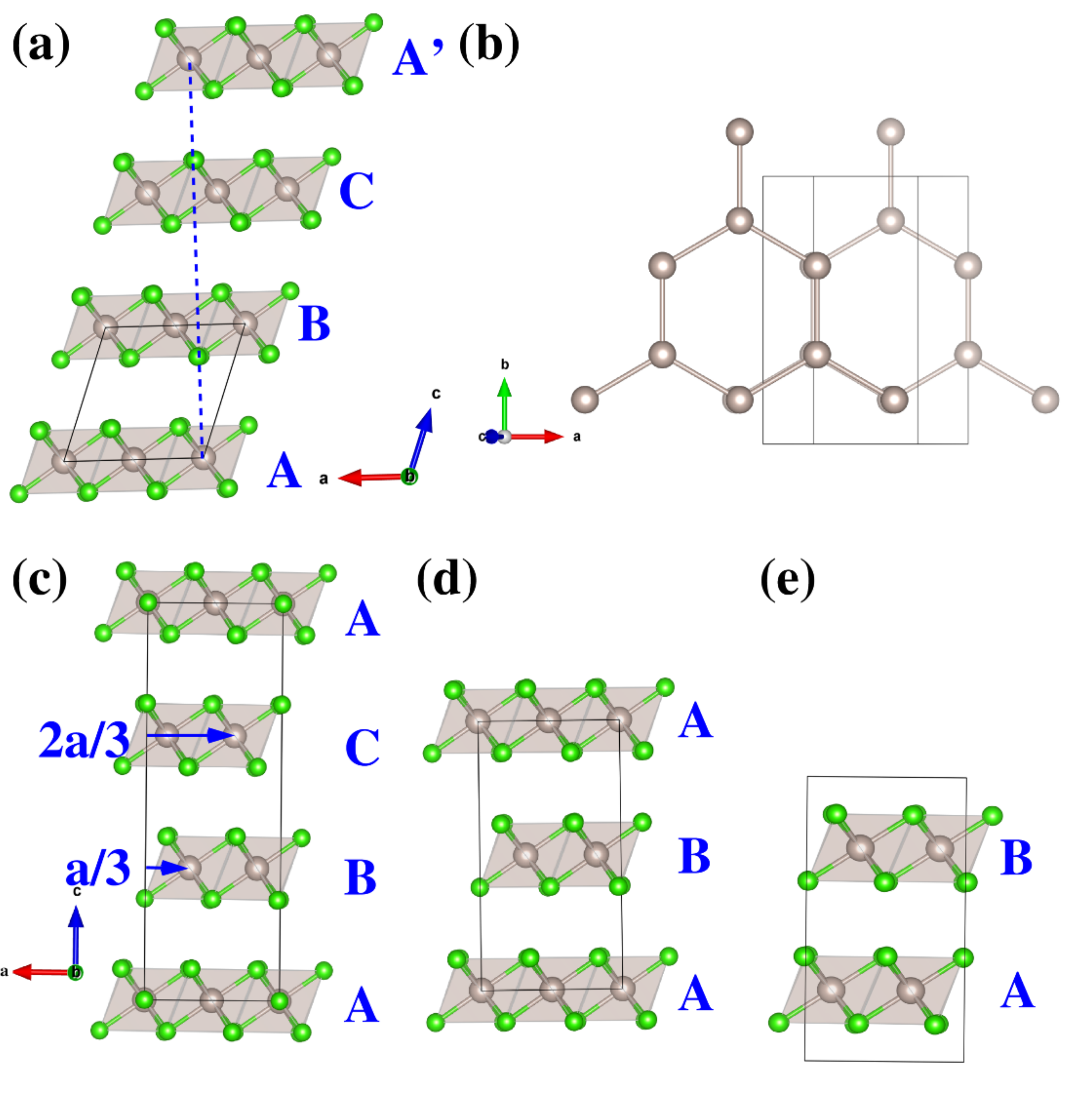}
\caption{ \label{plot:allstacking}
Structures and stacking orders proposed for \RuCl. Grey and green spheres represent Ru and Cl atoms, respectively. Panel (a) shows the $C2/m$ structure~\cite{Pollini1994,Johnson2015,Cao2016} with panel (b) highlighting that the A and A' layers almost overlap when viewed along the $c^\ast$ direction. The angle between direction $\protect\overline{\text{AA'}}$ and the $a$ axis is $\SI{89.523}{\degree}$. Panel (c) presents the ABC stacking proposed in Ref.~\cite{Cao2016} (see Fig. 1 (b) there), which is almost identical to the $C2/m$ structure since $a/3\approx -c\cos \beta$. Panel (d) shows the AB stacking proposed in Ref~\cite{Cao2016} (see Fig. 1 (c) there) with panel (e) representing a unit cell for this AB stacking within the $C2/m$ symmetry obtained by allowing bond length differences of up to $10^{-4}\SI{}{\angstrom}$.
}
\end{figure*}

\begin{figure*}[htbp] \center
\includegraphics[width=\textwidth, clip, trim= 4.0mm 10.0mm 5.0mm 7.0mm]{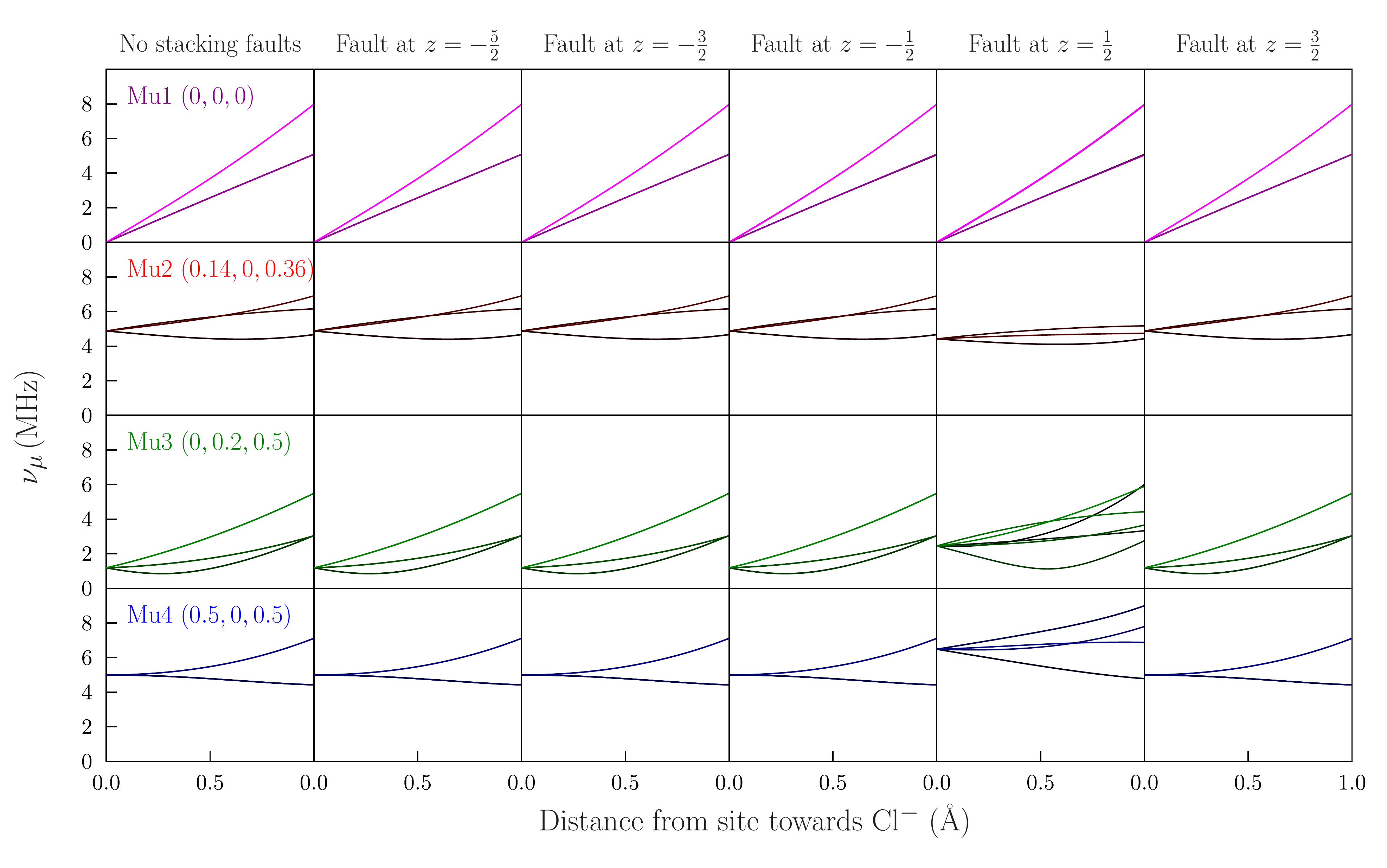}
\caption{ \label{plot:stackoakridge}
Muon precession frequencies as for Figure~\ref{plot:dipolestacking},
but now with ABAB-stacking order as shown in Fig. 1(c) of Ref.~\cite{Cao2016}.
}
\end{figure*}

An alternative stacking of the Ru layers in \RuCl has been proposed~\cite{Cao2016,Banerjee2016} and a previous neutron scattering study attributed the two transitions observed to two different stacking orders~\cite{Banerjee2016}. Figure~\ref{plot:allstacking} summarizes the different structures and stackings proposed for \RuCl. To investigate the feasibility of the AB stacking, which Banerjee {\it et al.}~\cite{Banerjee2016} highlight as the root for the higher temperature transition ($T_\text{N}\approx 14$~K) they observed, we performed DFT calculations for both the ABC and AB stacking orders shown in Figure~\ref{plot:allstacking} using the two pull potential all-electron codes WIEN2K~\cite{Wien2k} and FPLO~\cite{FPLO1,FPLO2}. 
For these calculations we prepare an AB unit cell out of the ABC unit cell (see Figure~\ref{plot:allstacking}) and keep the same atom positions without further structural relaxations. The total energy of the AB stacking turns out to be higher than the ABC stacking by a few tens of meV, making the AB stacking energetically less favorable.
Additionally, we confirmed that the electrostatic potential predicts the muon site candidates to be unchanged in the AB stacking  structure and calculated the dipole fields expected at these muon sites. The results, plotted in Figure~\ref{plot:stackoakridge}, show that the Larmor precession frequencies of the muons are essentially the same as those we calculated for the $C2/m$ structure. Therefore, it is unlikely that two different magnetic phases present in our sample due to the two proposed stacking orders (ABC and AB) is at the root of the two frequencies we observed in our \musr experiment.


\subsection{Alternative magnetic structure}
\begin{figure*}[htbp] \center
\includegraphics[width=\textwidth, clip, trim= 0.0mm 0.0mm 0.0mm 4.0mm]{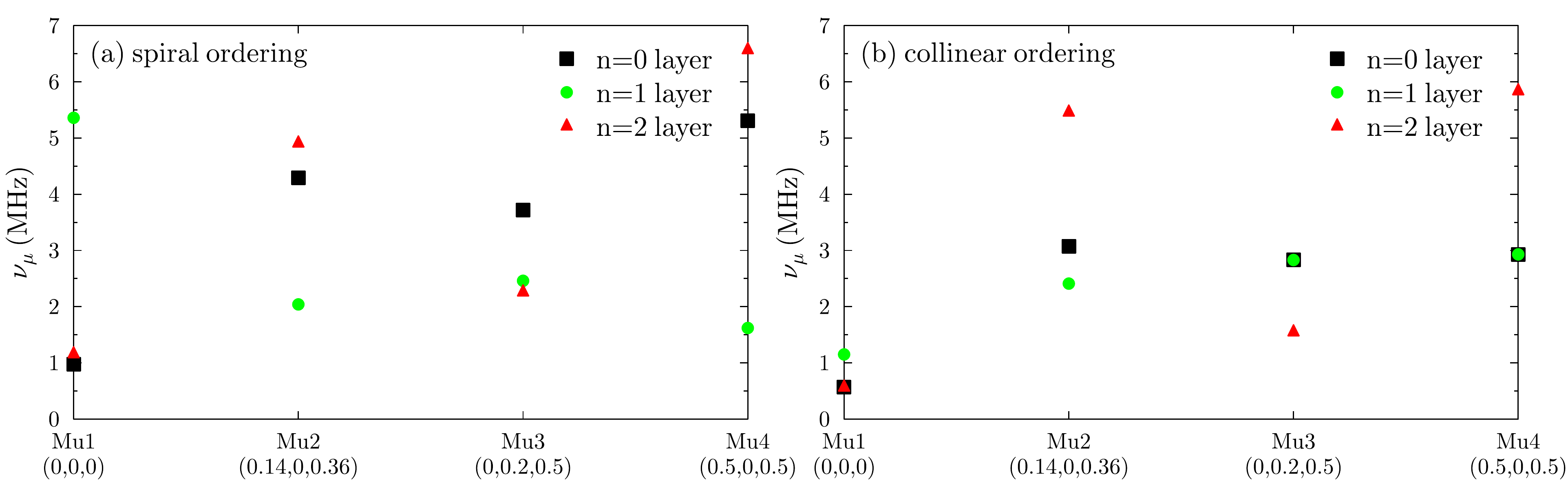}
\caption{ \label{plot:dipolecollinspiral}
Muon precession frequencies at the muon site candidates for the 3-layer magnetic orders proposed by Cao {\it et al}~\cite{Cao2016}. The parameter $n$ labels the Ru layer within the 3-layer stacking, see Figure 5 (a) and (b) in Ref.~\cite{Cao2016}.
}
\end{figure*}

A recent neutron diffraction study has proposed a magnetic structure in pristine single crystals of \RuCl with zig-zag antiferromagnetic ordering within the Ru layers but with 3-layer stacking periodicity~\cite{Cao2016}. Based on this magnetic structure we have calculated the dipole fields at the muon site candidates obtained from the earlier DFT analysis. Figure~\ref{plot:dipolecollinspiral} presents the resulting muon precession frequencies for the muon site candidates in each of the three distinct layers of both the proposed spiral and collinear spin configurations. We note that the magnitudes of the frequencies are of the order of the observed ones and that a more realistic analysis including distortions of the muon sites towards the nearest Cl$^-$ ions could potentially improve the quantitative agreement. However, the spiral and collinear ordering lead to three and two distinct frequencies, respectively, associated with muons stopping near the three different Ru layers. Additionally, including the more realistic distortions towards nearby Cl$^-$ ions will in general lift the degeneracies of the frequencies associated with Cl$^-$ ions that are symmetry equivalent in the crystallographic unit cell but are not equivalent in the larger magnetic unit cell. Therefore, the number of precession frequencies we expect to observe experimentally in such a magnetic structure is larger than two. Hence, the 3-layer magnetic structure proposed by Cao {\it et al.} for single crystals is incompatible with our \musr measurements of \RuCl powder. Nevertheless, another \musr measurement on pristine \RuCl single crystals is required to confirm whether the 3-layer ordering is indeed the appropriate one and might provide important insights into the proposed change of magnetic structure induced by mechanical deformations of single crystals.

\vspace{10mm}

\bibliography{aRuCl3}